\documentclass[epj,nopacs,twocolumn]{svjour} 
\usepackage{graphics}
\usepackage{graphicx} 
\usepackage{siunitx}
\usepackage{graphicx}  
\usepackage{amsmath, amssymb}  
\usepackage{hyperref}  
\usepackage{authblk}  
\usepackage{cite}  
\usepackage{lineno}
\usepackage{makecell}
\usepackage{subcaption}
\usepackage[normalem]{ulem}
\begin{document}
\title{Enhancing Electromagnetic Calorimeter Signal Reconstruction with Machine Learning-Based Noise Discrimination}

\author{
  Suman Das Gupta\inst{1}\fnmsep\thanks{\email{sumandg1997@gmail.com}} \and
  Shamik Ghosh\inst{2}\fnmsep\thanks{\email{shamik.ghosh@cern.ch (corresponding author)}} \and
  Laltu Gazi\inst{1}\fnmsep\thanks{\email{laltu.gazi@cern.ch}} \and
  Shubham Dutta\inst{3}\fnmsep\thanks{\email{shubhamdutta\_16@yahoo.com}} \and
  Alexander Ledovskoy\inst{4}\fnmsep\thanks{\email{alexander.ledovskoy@cern.ch}} \and
  Satyaki Bhattacharya\inst{1}\fnmsep\thanks{\email{bhattacharya.satyaki@saha.ac.in}} \and
  Shilpi Jain\inst{5}\fnmsep\thanks{\email{shilpi.jain@cern.ch}}
}

\institute{
  High Energy Nuclear and Particle Physics Division, Saha Institute of Nuclear Physics - a CI of Homi Bhabha National Institute, Kolkata - 700064, INDIA \and
  Laboratoire Leprince-Ringuet, CNRS/IN2P3, Ecole polytechnique, Institut Polytechnique de Paris, Palaiseau, France \and Department of Physics and Astronomy, Stony Brook University - SUNY, New York, United States \and
  University of Virginia, United States \and
  Department of High Energy Physics, Tata Institute of Fundamental Research, Mumbai, India  
}
\date{Received: date / Revised version: date}
%
\abstract{
Calorimeters operating in high-radiation environments are susceptible to damage, leading to increased noise that can significantly degrade energy resolution. A common way to mitigate noise is to apply a higher energy threshold on the calorimeter cells, typically set a few standard deviations above the noise level. However, this method risks discarding cells with genuine energy deposits, worsening the energy resolution and the energy deposit pattern. In this paper, we investigate graph neural network (GNN) based algorithms as an alternative to rigid energy thresholds. The proposed approach exploits the full pulse-shape information together with the correlations among energy deposits in individual cells within an electromagnetic cluster. The study is performed using a standalone Geant4 simulation of an $11\times11$ matrix of lead tungstate crystals. To simulate it in more relatic way We demonstrate that the machine learning based method significantly outperforms a simple threshold based strategy, achieving an improvement in the calorimeter energy resolution of up to $82\%$ and recovering up to $83\%$ of true signal cells within the calorimeter grid.
%
} 
\maketitle
\section{Introduction}
\label{intro}
High-energy physics experiments, such as those conducted at the Large Hadron Collider (LHC)~\cite{Evans:2008zzb}, rely on calorimeters for precise energy measurement of particles, which play a crucial role in reconstructing electrons, photons, jets, and missing transverse energy (MET). However, in the high-radiation environment of the LHC, calorimeters are subjected to continuous exposure to intense particle fluxes, leading to radiation-induced damage over time. This degradation manifests as increased noise levels relative to signal amplitude, leading to reduced signal integrity and worsening of the energy resolution~\cite{CMS:2024ppo}. This directly impacts the accuracy of reconstructed physics objects. As the total integrated luminosity increases, the accumulation of radiation damage progressively deteriorates the detector performance, affecting the sensitivity of the experiment to rare or subtle physics processes.

A widely used method to suppress noise in calorimeters is the application of a fixed energy threshold on individual calorimeter cells. Typically, this threshold is set a few standard deviations above the measured noise level, ensuring that spurious signals are minimized. However, such an approach comes with significant limitations: it does not differentiate between real low-energy electromagnetic (EM) hits and noise, leading to potential loss of genuine physics signals~\cite{CMS:2024ppo}. 
For instance, in scenarios where EM showers deposit energy across multiple calorimeter cells, a rigid thresholding scheme may discard low-energy contributions, thereby underestimating the total energy of the object. This worsens the energy resolution, affecting key physics analyses such as precision measurements of Higgs boson mass, searches for new particles, and detailed studies of jet substructure.

Given these challenges, machine learning (ML) techniques offer a promising alternative by leveraging the full temporal information of the pulse shape~\cite{Dutta_2023} including out-of-time (OOT) pileup along with spatial and energy information of calorimeter signals. Unlike conventional thresholding, ML-based approaches can learn complex correlations in responses of individual calorimeter cells, allowing for more powerful  discrimination between genuine EM energy deposits and noise-induced fluctuations. Furthermore, ML models can be trained to adapt dynamically to evolving detector conditions, making them particularly suitable for long-term operation in radiation-intensive environments.

In this paper, we explore the application of various ML algorithms for noise suppression in a homogeneous electromagnetic calorimeter, focusing on their ability to enhance energy reconstruction, improve resolution, and ultimately extend the physics reach of collider experiments. We compare different ML architectures, and evaluate their performance on realistic calorimeter data. Our results demonstrate that ML-based noise discrimination can significantly reduce energy bias, improve resolution, and enhance the overall robustness of calorimeter-based measurements, offering a data-driven solution for mitigating radiation-induced detector degradation.

\section{Geant4 Simulation and Monte Carlo Generation}
\label{sec:sim}
The model is trained on samples generated using the Geant4-based~\cite{GEANT4} calorimeter simulation. We use a simple geometry of an $11 \times 11$ array of PbWO$_4$ crystals, each measuring $2.2 \times 2.2\,\mathrm{cm}^2$ in cross section and $22\,\mathrm{cm}$ in length, with the interaction vertex located 151 cm upstream of the front face to emulate the geometry to the barrel section of the EM calorimeteter of the Compact Muon Solenoid (CMS)~\cite{CMS:2008xjf} experiment at the LHC, CERN. 

\begin{figure}[htbp]
\centering
\includegraphics[width=0.48\textwidth]{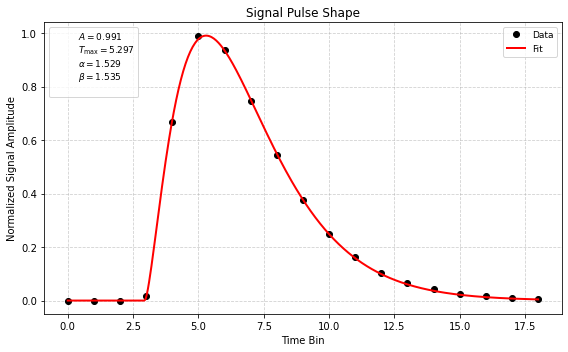}
\caption{Pulse shape binned template. The x-axis is given in units of 25 ns. The first 3 bins
are the pedestal samples, and their values equal zero by construction. The following 15 bins are estimated from the Equation~\ref{pulseshape} with parameters shown in the legend box.}
\label{fig:Signal_Pulse_Template}
\end{figure}

To simulate the signal pulse for each crystal, mono energetic electrons are fired perpendicularly onto the front face of the crystal array. In each simulated event, the energy deposited in the $i^{th}$ crystal is recorded as $E^{\,i}_{\mathrm{cell}}$. The digitization procedure is implemented in accordance with the methodology adopted in the current CMS electromagnetic calorimeter (ECAL) \cite{CMS:1997ysd}. 
To emulate the digitization process in the simulation, the deposited energy $E^{\,i}_{\mathrm{cell}}$ is used to scale a normalized pulse template, $S(t)$, which is parameterized as described in Equation~\ref{pulseshape}~\cite{CMS:2020xlg}. This pulse shape template is then engineered to include three presamples, which are essential for performing pedestal subtraction. The final pulse template is sampled at a frequency of 40\,MHz, corresponding to the bunch crossing rate of the LHC, and comprises 19 discrete time samples. The pulse is timed such that its peak occurs at the sixth sample. Figure~\ref{fig:Signal_Pulse_Template} shows an example of the signal pulse shape template. The characteristics of the simulated waveform, including its rise time and sampling structure, have been selected to faithfully reproduce the known response of the CMS ECAL~\cite{CMS:2024ppo}.


\begin{equation}
S(t) =
\begin{cases}
0, & \Delta t < -\alpha\,\beta, \\[8pt]
\left( 1 + \dfrac{\Delta t}{\alpha\,\beta} \right)^{\alpha}
\exp\!\left( -\dfrac{\Delta t}{\beta} \right), & \Delta t \ge -\alpha\,\beta,
\end{cases}
\label{pulseshape}
\end{equation}
where $\Delta t = t - T_{\mathrm{max}}$ represents the time position relative to the peak, and $\alpha, \beta$ are two shape parameters.

To model electronic noise, a zero pedestal is assumed, and an RMS noise amplitude of $ \sigma_{\mathrm{noise}} = 150$~MeV is applied. First, uncorrelated noise samples $n_{\mathrm{uncorr}}(t) \sim \mathcal{N}(0, \sigma)$ are generated independently for each time index $t = 0 \dots 18$. These raw samples are then transformed to include 
inter- sample correlations by forming a covariance matrix $\rho_{t_1, t_2}$ between the time samples that follows an exponential law whose time constant is related to the shaping time of the electronics as discussed in~\cite{Adzic:2006hda}. 
A Cholesky decomposition of this covariance matrix is used to convert $\{ n_{\mathrm{uncorr}}(t) \}$ into a correlated noise vector $\{ N^{\,i}(t) \}$ for each crystal. 

In order to include the effect of additional pp collisions, i.e. realistic pileup conditions, we simulate minimum bias (minbias) events (pp collisions) using \texttt{Pythia8} (v8.307)~\cite{pythia8} at $\sqrt{s} = 13.6$ TeV corresponding to the Run3 LHC energy. The beam configuration is rotated such that the event energy flow aligns with pseudorapidity $\eta$. Events are stored in HepMC3 format~\cite{BUCKLEY2021107310} and propagated through the geometry, recording the energy deposited in each crystal $i$, $E^{\,i}_{\mathrm{minbias}}$, for a single minimum‑bias interaction. 


In this study, we consider the effect of 15 past bunch crossings (BX), the in-time BX, and 9 post BXs. This selection is based on the final pulse template, which employs only 10 sampling points, making effects after 10 BX irrelevant. Furthermore, tail contributions beyond 15 past BXs are negligible and do not affect the final pulse shape. To simulate LHC Run~3 pileup conditions, the number of overlapping minbias interactions per BX is drawn from a Poisson distribution with mean $\langle \mu \rangle = 50$ consistent with Run~3 conditions at the LHC. For each of the twenty-five BX indices (BX=$-15 \dots -1$, $0$, $+1 \dots +9$), an independent Poisson count $N_{\mathrm{minbias}}$ is sampled; $N_{\mathrm{minbias}}$ random deposits $E^{\,i}_{\mathrm{minbias}}$ are summed to compute $E^{\,i}_{\mathrm{pileup},\,j}$ in crystal $i$ for BX index $j$. 

\begin{figure*}[htbp]
\centering
\includegraphics[width=0.48\textwidth]{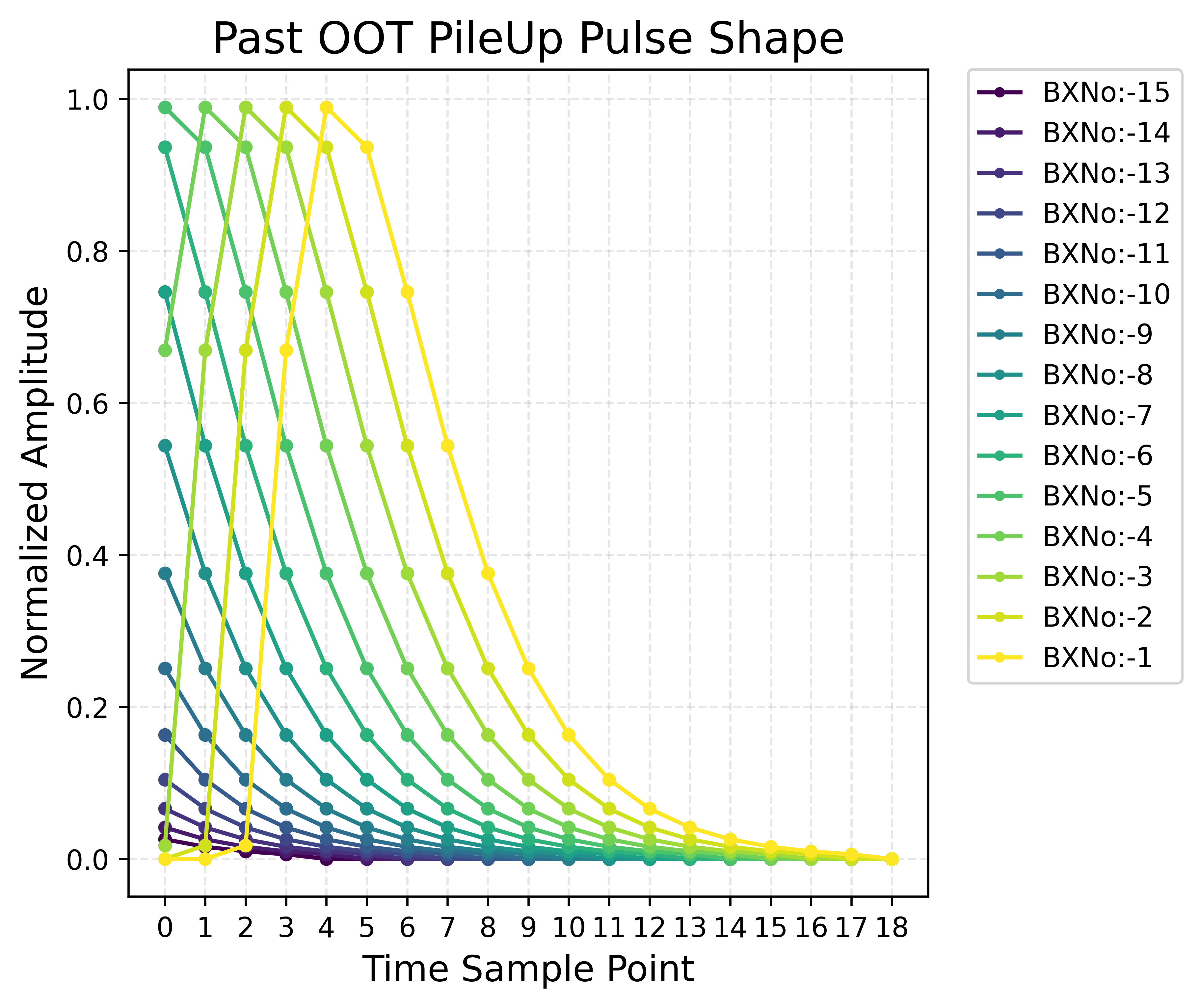}
\hfill 
\includegraphics[width=0.48\textwidth]{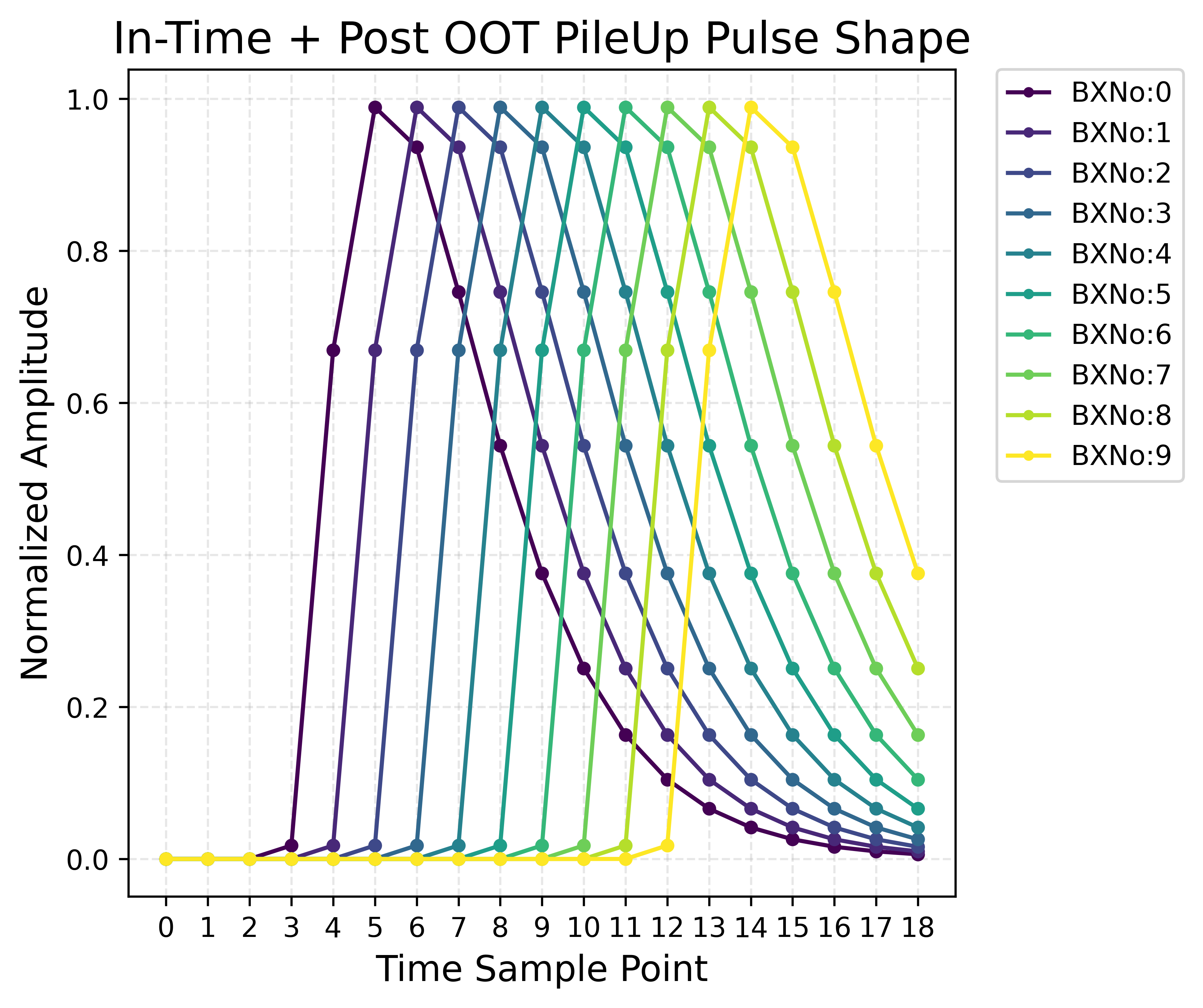}
\caption{Pileup pulse templates for different bunch crossings (BX). Each template is constructed by time-shifting \(S(t)\) by integer multiples of the BX spacing. The left panel shows the 15 preceding BX templates; the right panel shows the in-time BX and the 9 subsequent BX templates.}
\label{fig:pileup_pulse_template}
\end{figure*}

The analog response of each deposited energy is modeled by $S(t)$. Twenty five shifted templates $\{ S_{-15}, \dots, S_0, \dots, S_{+9} \}$ are constructed by translating $S(t)$ in time by integer multiples of the BX spacing. These pileup pulse templates are shown in Figure~\ref{fig:pileup_pulse_template}. For BX index $j$, the template $S_j(t)$ is scaled by $E^{\,i}_{\mathrm{pileup},\,j}$; summing over all BXs produces the total pileup waveform:
\[
P^{\,i}_{\mathrm{pileup}}(t) = \sum_{j = -15}^{+9} E^{\,i}_{\mathrm{pileup},\,j} \, S_j(t).
\]

Finally, for each crystal, the final time‑sample sequence is:

\begin{equation}
T^{\,i}(t) = E^{\,i}_{\mathrm{signal}}\, S_0(t) + \sum_{j = -15}^{+9} E^{\,i}_{\mathrm{pileup},\,j}\, S_j(t) + N^{\,i}(t),
\label{eqn:totalE}
\end{equation}
providing realistic ADC readout samples that include contributions from signal, pileup (past, in‑time, and post BXs), and noise.
Here, $S_{0}$ and $E^{\,i}_{\mathrm{signal}}$ correspond to the pulse shape and the energy deposited in $i^{th}$ crystal from the current BX, while $S_j(t)$ and $E^{\,i}_{\mathrm{pileup},\,j}$ represent the contributions from pileup interactions in the $j^{th}$ BX. The term $N^{\,i}(t)$ denotes the noise waveform for the $i^{th}$ crystal. The final pulse time-sample is constrained to 10 sampling points, instead of the simulated 19, to align with the ECAL digitizer time window.

Following the above prescription, 10k events were generated  for each photon energy ranging from \SI{5}{\giga\electronvolt} to \SI{30}{\giga\electronvolt} in steps of \SI{5}{\giga\electronvolt} at pseudorapidity \(\eta = 1.4\). The choice of \(\eta = 1.4\) targets a high-pileup region within the CMS barrel, representing a challenging scenario for reconstruction. 

\section{Cell Energy Reconstruction}

To mitigate the impact of pileup from both past and future BXs, we employ the pulse-shape fitting technique used in CMS~\cite{CMS:2020xlg}, commonly referred to as the multifit algorithm.
The multifit algorithm aims to reconstruct the energy deposited \( E_{\mathrm{reco}}^i \) in a given crystal by fitting a model to the observed time-sample sequence \( \boldsymbol{T}^i(t) \).

The algorithm models the expected waveform as a linear combination of known pulse templates from the signal and pileup contributions across multiple BXs, optimizing the amplitudes \( A_j^{(i)} \) to best match the observed data.

The fit minimizes the \( \chi^2 \) defined as:
\begin{equation}
\chi^{2}_{i} =
\biggl( \sum_{j=0}^{N_{\mathrm{BX}}} A_{j}^{(i)} \boldsymbol{S}_{j} - \boldsymbol{T}^{i} \biggr)^{\!T}
\mathbf{C}_{i}^{-1}
\biggl( \sum_{j=0}^{N_{\mathrm{BX}}} A_{j}^{(i)} \boldsymbol{S}_{j} - \boldsymbol{T}^{i} \biggr),
\end{equation}
subject to \( A_{j}^{(i)} \geq 0 \) for all \( j \).

\begin{figure}[htbp]
\centering
\includegraphics[width=0.48\textwidth]{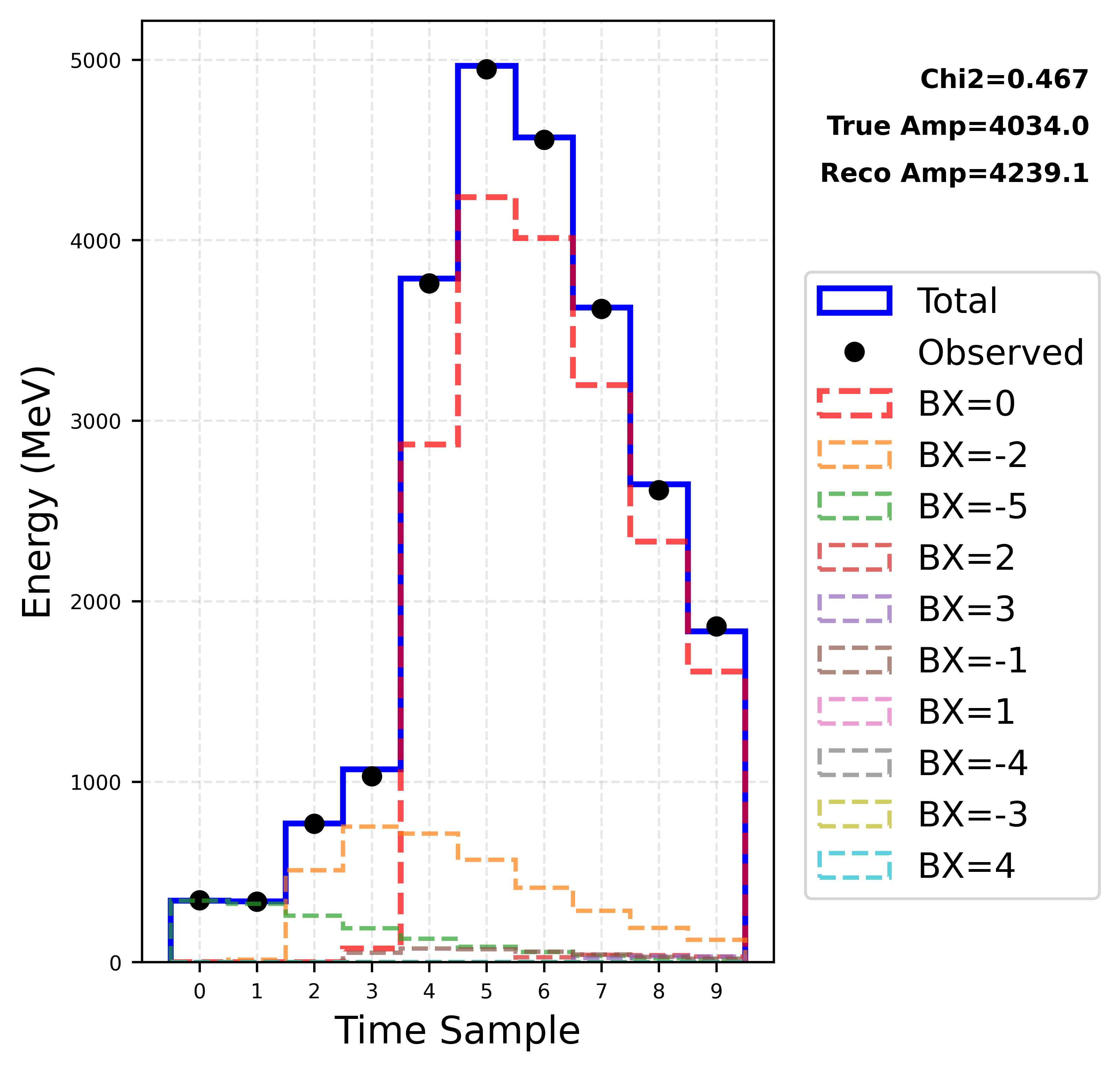}
\caption{Example of a fitted pulse using the multifit algorithm in a single crystal. Filled circles represent the 10 digitized samples, the red dashed distributions (dotted multicolored distributions) represent the fitted in-time (out-of-time) pulses with positive amplitudes. The solid dark blue histograms represent the sum of all the fitted contributions.}
\label{fig:Mulifit_Performance}
\end{figure}

\begin{figure}[htbp]
\centering
\includegraphics[width=0.44\textwidth]{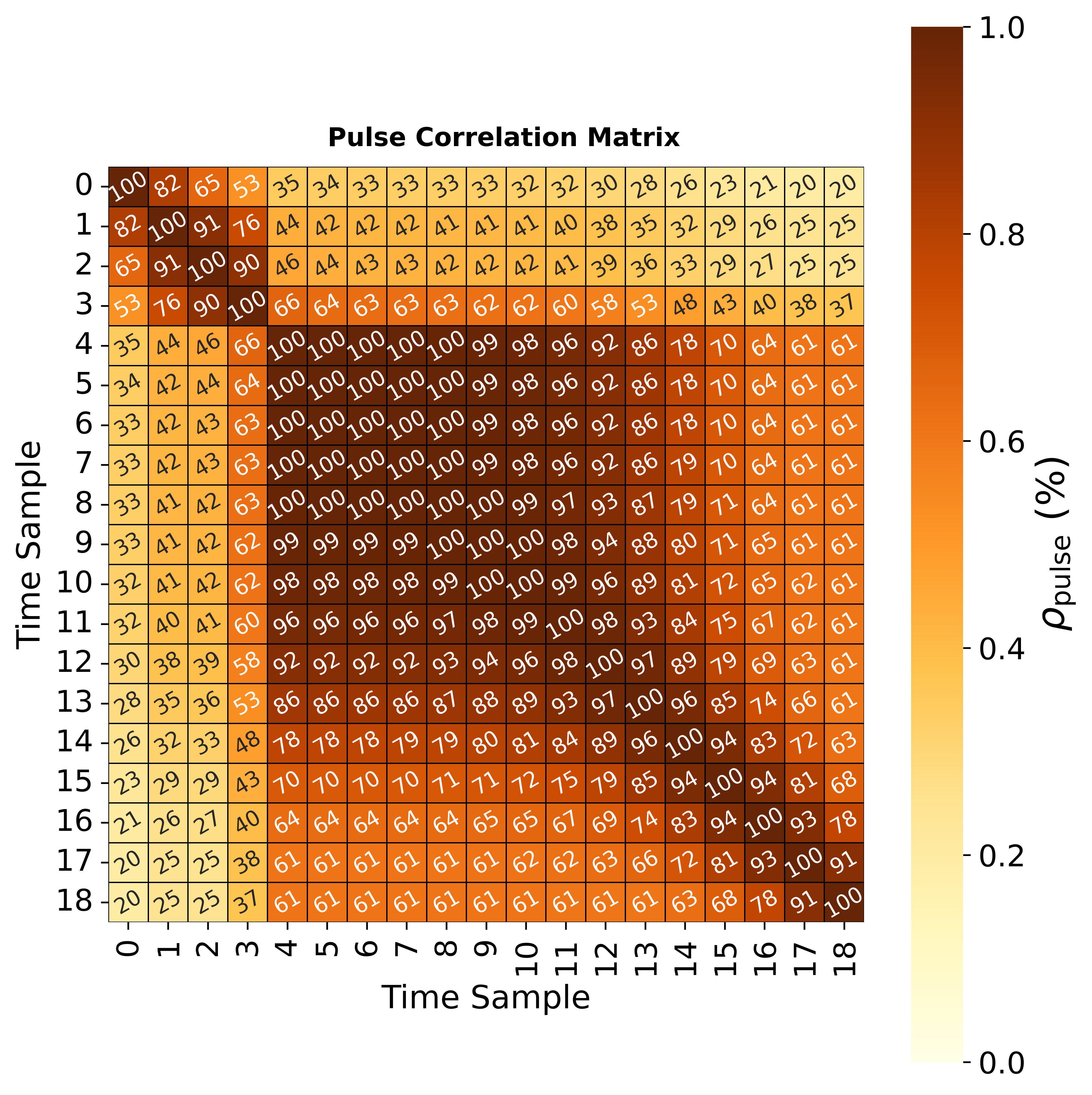}
\includegraphics[width=0.44\textwidth]{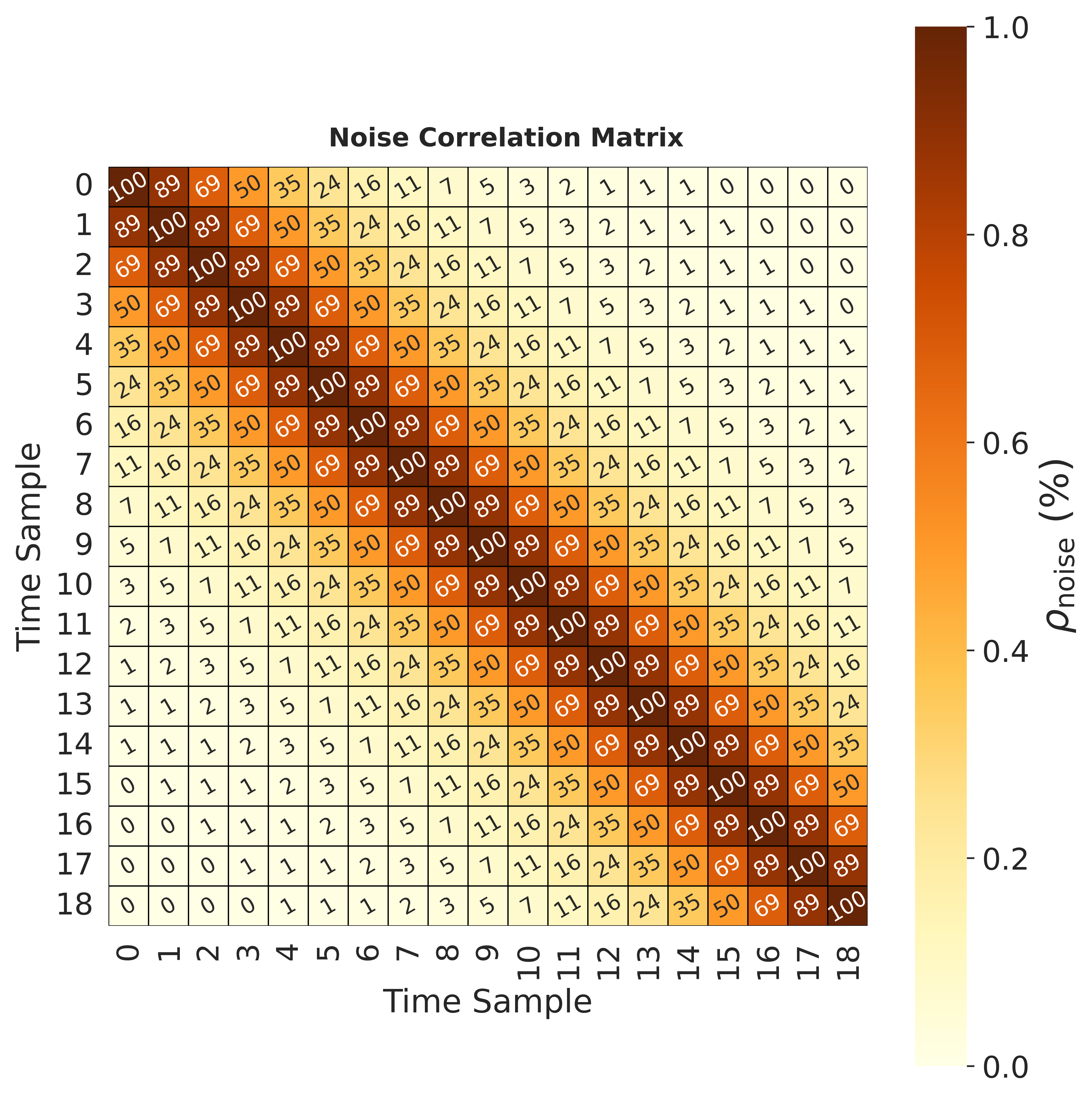}
\caption{Correlation matrices of the pulse shape binned templates (top) and the electronics noise (bottom). All the
values in the figure are expressed in percent for legibility}
\label{fig:correlation_matrix}
\end{figure}

This is solved via a nonnegative least-squares (NNLS) solver.  
The process initializes \( A_j^{(i)} = 0 \), iteratively adds nonzero amplitudes, updates \( \mathbf{C}_i^{-1} \) in the active template subspace, and stops when \( \Delta \chi^2 < 10^{-3} \) or no further improvement occurs.  
The in-time amplitude \( A_0^{(i)} \) yields the cell energy:
\[
E_{\mathrm{reco}}^{(i)} = A_0^{(i)}.
\]
Example of a fitted pulse using the multifit algorithm in a single crystal is shown in Figure~\ref{fig:Mulifit_Performance}.
Here, \( \mathbf{C}_{i} \) is the full \( 10 \times 10 \) covariance matrix of the samples in crystal \( i \), defined by:
\[
\mathbf{C}^i = \mathbf{C}_{\mathrm{noise}}^{(i)} + \sum_{j} (A_j^{(i)})^2 \mathbf{C}_{\mathrm{pulse}}^{(i)},
\]
with
\[
\mathbf{C}_{\mathrm{noise}}^{(i)} = \sigma_{\mathrm{noise}}^2 \, \rho_{\mathrm{noise}}(t_1, t_2),
\]
where $\rho_{\mathrm{noise}}(t_1, t_2)$ is constructed using as discussed in Section~\ref{sec:sim}, and $\sigma_{\mathrm{noise}}$ is set to 150 $\mathrm{MeV}$. The correlation matrix of the noise is shown in Figure~\ref{fig:correlation_matrix} (bottom).

The matrix \( \mathbf{C}_{\mathrm{pulse}}^{(i)} \) is computed from \( N_{\mathrm{ev}} \) simulated signal pulses with signal energy deposition \( \mathbf{E}_{\mathrm{true}}^i > 1~\mathrm{GeV} \):

\[
\bigl[ \mathbf{C}_{\mathrm{pulse}}^{(i)} \bigr]_{t_1, t_2} = \frac{1}{N_{\mathrm{ev}}} \sum_{n=1}^{N_{\mathrm{ev}}} \widetilde{T}^i(t_1, n) \, \widetilde{T}^i(t_2, n),
\]
where \( \widetilde{T}^{i}(t, n) \) is the normalized sample \( t \) in event \( n \), scaled so that \( \widetilde{T}_{i}(t_{\mathrm{peak}}, n) = 1 \). 
The correlation matrix, \( \rho_{\mathrm{pulse}} \), of the pulse template is shown in Figure~\ref{fig:correlation_matrix} (top), is defined as
\[
\rho^{i,k}_{\mathrm{pulse}} = \frac{C^{i,k}_{\mathrm{pulse}}}{\sigma^{i}_{\mathrm{pulse}} \, \sigma^{k}_{\mathrm{pulse}}}\,,
\]
where \( \sigma^{i}_{\mathrm{pulse}} \) and \( \sigma^{k}_{\mathrm{pulse}} \) denote the square roots of the variances of the pulse shape in bins \( i \) and \( k \) of the template, respectively.

\section{Methodology for Graph-Based Analysis}



To ensure robust training of the machine learning models, a balanced dataset comprising an adequate number of signal and noise cells is essential to mitigate data imbalance. As photon energy increases, the electromagnetic shower broadens, reducing the fraction of noise cells within an $11 \times 11$ grid, particularly at higher energies. To address this, we optimize the minimum energy threshold for classifying cells as signal (above the threshold) or noise (below the threshold), targeting a balanced dataset for photon energies from \SI{5}{\giga\electronvolt} to \SI{30}{\giga\electronvolt}. This threshold is carefully selected to minimize its impact on the mean photon energy and energy resolution.

\begin{figure*}[htbp]
\centering
\includegraphics[width=0.32\textwidth]{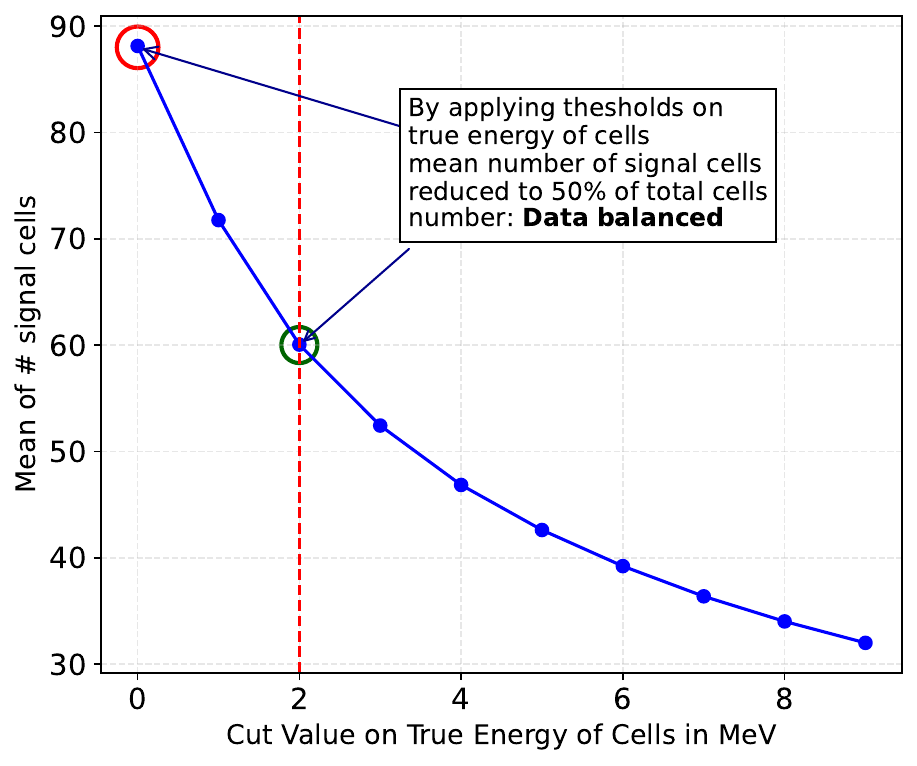}
\hfill 
\includegraphics[width=0.32\textwidth]{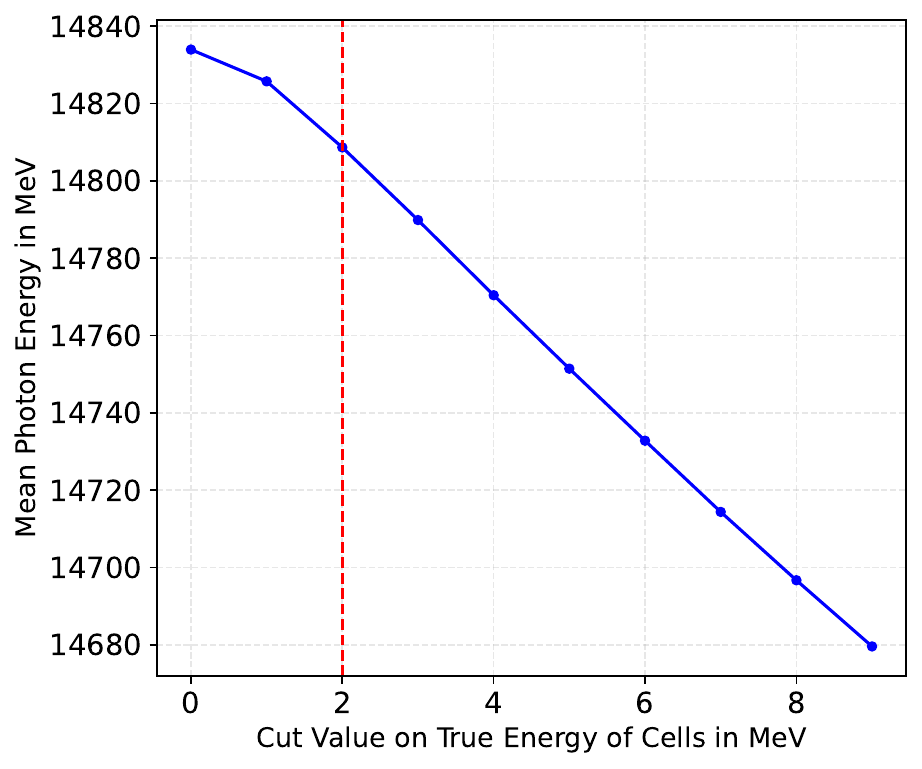}
\hfill 
\includegraphics[width=0.32\textwidth]{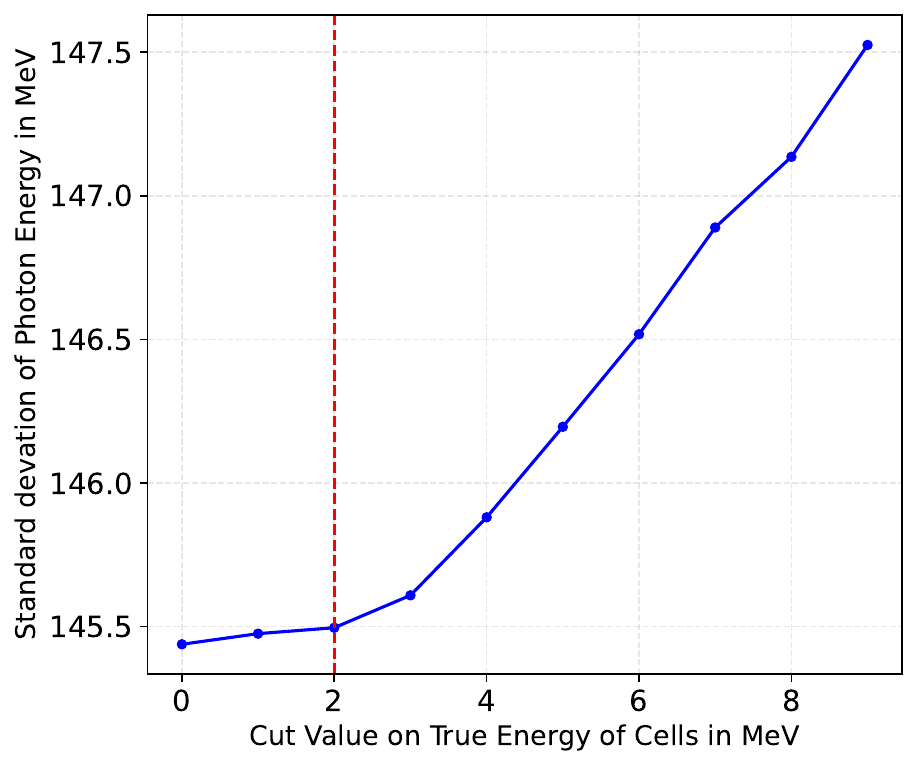}

\caption{Variation of the fraction of true signal cells (left), mean of photon energy (middle), and standard deviation of photon energy (right) as a function of the energy threshold on \(E_{\text{cell,true}}\) for 15 GeV photons.}
\label{fig:photon_metrics_CutOpt}
\end{figure*}

We systematically evaluated the effects of varying the minimum energy threshold on key performance metrics, including the mean photon energy, the standard deviation of photon energy, and the fraction of signal cells. The results are presented in Figure~\ref{fig:photon_metrics_CutOpt}, which illustrates the variation of these metrics as a function of the energy threshold for \SI{15}{\giga\electronvolt} photons. As shown in Figure~\ref{fig:photon_metrics_CutOpt} (left), a threshold of $2\,\text{MeV}$ reduces the mean number of signal cells to $50\%$ of the total cells in the grid. Concurrently, the impact on the mean photon energy (Figure~\ref{fig:photon_metrics_CutOpt}, middle) and the standard deviation of photon energy (Figure~\ref{fig:photon_metrics_CutOpt}, right) is minimal. Quantitative results are summarized in Table~\ref{tab:photon_metrics_CutOpt}, detailing the impact of the chosen threshold on the mean photon energy and the standard deviation of photon energy compared to a scenario without any energy cut on true cell energy (\(E_{\text{cell,true}}\)). The findings indicate that the optimized threshold effectively balances the dataset while preserving energy resolution and mean photon energy, thereby enhancing the ML model's performance for high-energy photon detection.

\begin{table*}[htbp]
\centering
\caption{Impact of the optimized threshold on mean photon energy and energy resolution compared to the scenario without any energy cut on true cell energy.}
\label{tab:photon_metrics_CutOpt}
\begin{tabular}{lcccc}
\hline\noalign{\smallskip}
Photon Energy & $E_{\text{th}}$ Cut & Rel. Loss in & Rel. Loss in & Mean \# of \\
(GeV) & (MeV) & Mean Energy & Std. Energy & Signal Cells \\
\noalign{\smallskip}\hline\noalign{\smallskip}
5  & 0 & 0\%       & 0\%       & 63.55 \\
10 & 1 & $0.084$\%  & 0.090\%   & 62.53 \\
15 & 2 & $0.170$\%   & 0.040\%   & 60.04 \\
20 & 3 & $0.226$\%  & 0.053\%   & 58.61 \\
25 & 4 & $0.267$\%  & 0.018\%  & 57.61 \\
30 & 5 & $0.299$\%  & 0.151\%   & 56.84 \\
\noalign{\smallskip}\hline
\end{tabular}
\end{table*}

To ensure uniformity across events, the pulse shapes were max-normalized by dividing each sample by the maximum value of the waveform, such that the peak amplitude is unity. The resulting dataset was randomly partitioned into 75\% for training and 25\% for testing, providing a balanced split for model development and evaluation.

\subsection{Graph Construction}
For each event, a graph was constructed with nodes representing calorimeter cells. Node features include:
\begin{itemize}
    \item Pulse-shape: 10 time-sample points.
    \item Position of the cell: \((X^i_{\text{cell}}, Y^i_{\text{cell}})\).
    \item Normalized reconstructed energy,      \(E^i_{\text{N}}\): \(\frac{E^i_{\text{reco}}}{E^{\text{seed}}_{\text{reco}}}\), representing the cell’s energy relative to the cell with the highest energy deposit referred to as the seed.
\end{itemize}
Edges were defined based on spatial proximity, using the k-Nearest Neighbors~\cite{Hastie:2009esl} (\(k=24\)).

\section{Network Architecture}
\label{sec:netarch}
\begin{figure*}[htbp] 
    \centering
    \includegraphics[width=0.65\textwidth]{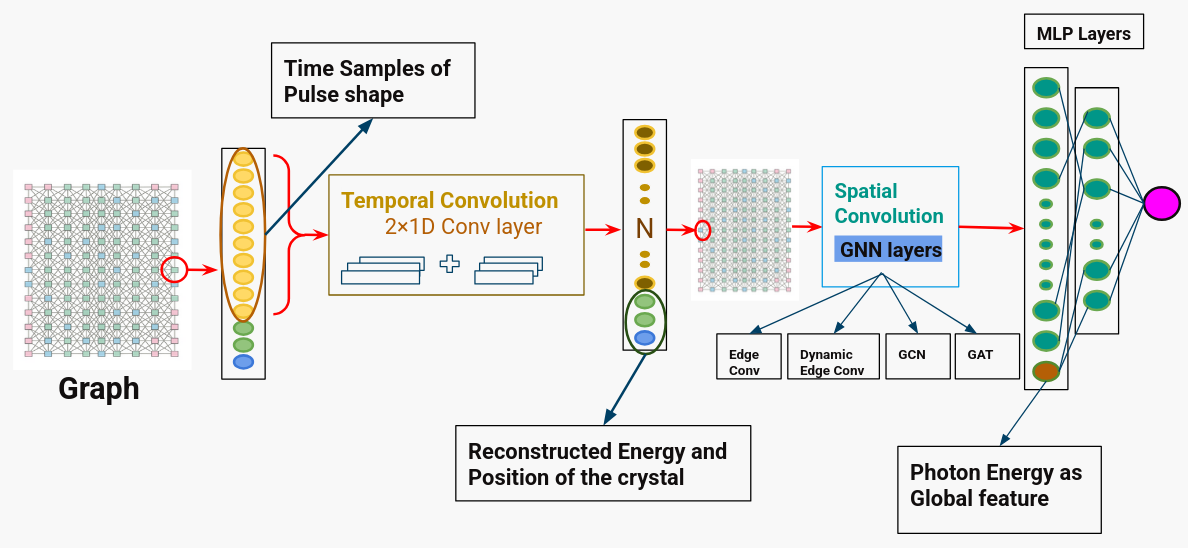}
    \caption{Flow chart of the model architecture}
    \label{fig:model_arch}
\end{figure*}

We frame the problem as a \textbf{node binary classification} task: each cell must be labeled as ``signal‐hit'' or ``noisy-hit.'' The network operates in three compact stages: (1) temporal feature extraction via one-dimensional convolutions~\cite{Krizhevsky:2012alexnet}, (2) graph‐based spatial aggregation using graph neural network (GNN)~\cite{Wu:2021gnn} layers, and (3) final per‐node classification. An overview of the data flow is shown in Figure~\ref{fig:model_arch}.

Each calorimeter cell is first processed by a \textbf{temporal encoder} that projects its 10-sample waveform into a fixed-length feature vector $\mathbf{h}^i$ with a dimensionality ranging from 16 to 64. Specifically, we apply two successive 1D convolutional layers (kernel size 3 and 2) with LeakyReLU activations, followed by a fully connected layer.  This encoder outputs $\mathbf{h}^i$, which serves as an extended descriptor of the local pulse shape for node $i$. 

The \textbf{node features} are then formed by concatenating these temporal embeddings with the cell’s spatial coordinates $(X^i, Y^i)$ and its reconstructed (normalized) energy \(E^i_{\text{N}}\), thereby combining temporal, spatial, and energy information for each node:
\[
\mathbf{x}^i = [\,\mathbf{h}^i \;\Vert\; X^i \;\Vert\; Y^i \;\Vert\; E^i_{\mathrm{N}}\,]
\]

We then apply a multi-layer GNN to propagate information over the constructed graph. We implement several graph convolution backbones, including a standard Graph Convolutional Network (GCN)~\cite{Kipf:2017gcn}, a Graph Attention Network (GAT)~\cite{Velickovic:2018gat}, Edge Convolution (Edge Conv), and Dynamic Edge Convolution (Dynamic Edge Conv)~\cite{Wang:2019dgcnn}. In each of these networks, we follow the same architecture. We stack multiple graph-convolution layers (typically 3 layers with 32-64 hidden units each and LeakyReLU activations), each optionally followed by dropout (typical dropout rate $\sim$0.15) to regularize the network. These layers learn node embeddings that capture the local structure in the graph. Following the final GNN layer, the photon energy is concatenated as a global feature with the learned node embeddings. For the GAT network, we define edge attributes as the dot product of the pulse shapes of connected nodes

Then, we apply a \textbf{readout classifier} to each node. The readout is a multi-layer perceptron (MLP) that takes the final node embedding and outputs a scalar probability of the cell being a signal hit. Concretely, we use two fully connected layers (e.g., hidden sizes 32 and 16) with LeakyReLU activations, followed by a sigmoid activation on the output to produce a score in $[0,1]$. Thus, the network performs binary classification at each node.

The model is trained with the \textbf{binary cross-entropy (BCE) loss}, \textbf{$\mathcal{L}_{\mathrm{BCE}}$}, for signal-hit/noise-hit classification. In addition to the standard BCE loss, two regularization terms are included to improve the physics performance. The first term minimizes the standard deviation (std) of the relative error on the reconstructed photon energy, calculated over each training mini-batch, which improves the energy resolution and encourages consistency in energy estimation using the Multifit algorithm. The second is a seed penalty term, \textbf{$\mathcal{L}_{\mathrm{seed}}$}, which ensures that the hit probability for the central crystal of the cluster, where the energy deposit is highest, is close to unity. The total loss $\mathcal{L}$ is given as:
\begin{equation}
\mathcal{L} = \mathcal{L}_{\mathrm{BCE}} + \mathrm{std}\left(\frac{E_{\mathrm{reco,photon}} - E_{\mathrm{true,photon}}}{E_{\mathrm{true,photon}}}\right) + \lambda \mathcal{L}_{\mathrm{seed}}
\end{equation}
where $E_{\mathrm{reco,photon}}$ is the total reconstructed photon energy in an event, and $E_{\mathrm{true,photon}}$ is the corresponding true photon energy. The reconstructed energy is calculated by weighting the energy of each hit, $E^{i}_{\mathrm{reco}}$, obtained from the Multifit algorithm, by the network's output signal probability, $p^i$:
\begin{equation}
E_{\mathrm{reco,photon}} = \sum_i p^i E^{i}_{\mathrm{reco}}
\label{eq:photon_energy_reco}
\end{equation}
The hyperparameter $\lambda$ is set to a large value of 100 to give substantial weight to the seed penalty term, $\mathcal{L}_{\mathrm{seed}}$. The network is trained using the Adam optimizer with a learning rate of $10^{-3}$ over multiple epochs, processing the data in mini-batches of graphs.

\section{Performance study}

This section presents the performance of the ML-based model and compares it with that of the traditional energy cut-based algorithm. In the first step, various GNN architectures, detailed in Section~\ref{sec:netarch}, are trained and tested on identical signal and noise datasets. The models are assessed using three primary metrics. The first is the area under the Receiver Operating Characteristic (ROC) curve (AUC) to evaluate signal-noise classification. The
second is the mean relative response of the weighted reconstructed photon energy,
as defined in Equation~\ref{eq:photon_energy_reco}, which evaluates the accuracy
of the energy scale. Finally, to complement the mean response, the third metric is the standard deviation of the relative error on the reconstructed photon energy with respect to the true energy, which quantifies the photon energy resolution improvement. This is the same quantity used as a regularization term in the loss function, but for this performance evaluation, it is calculated across all events for different ranges of true photon energy.

\begin{figure*}[htbp]
\centering

\includegraphics[width=0.32\textwidth]{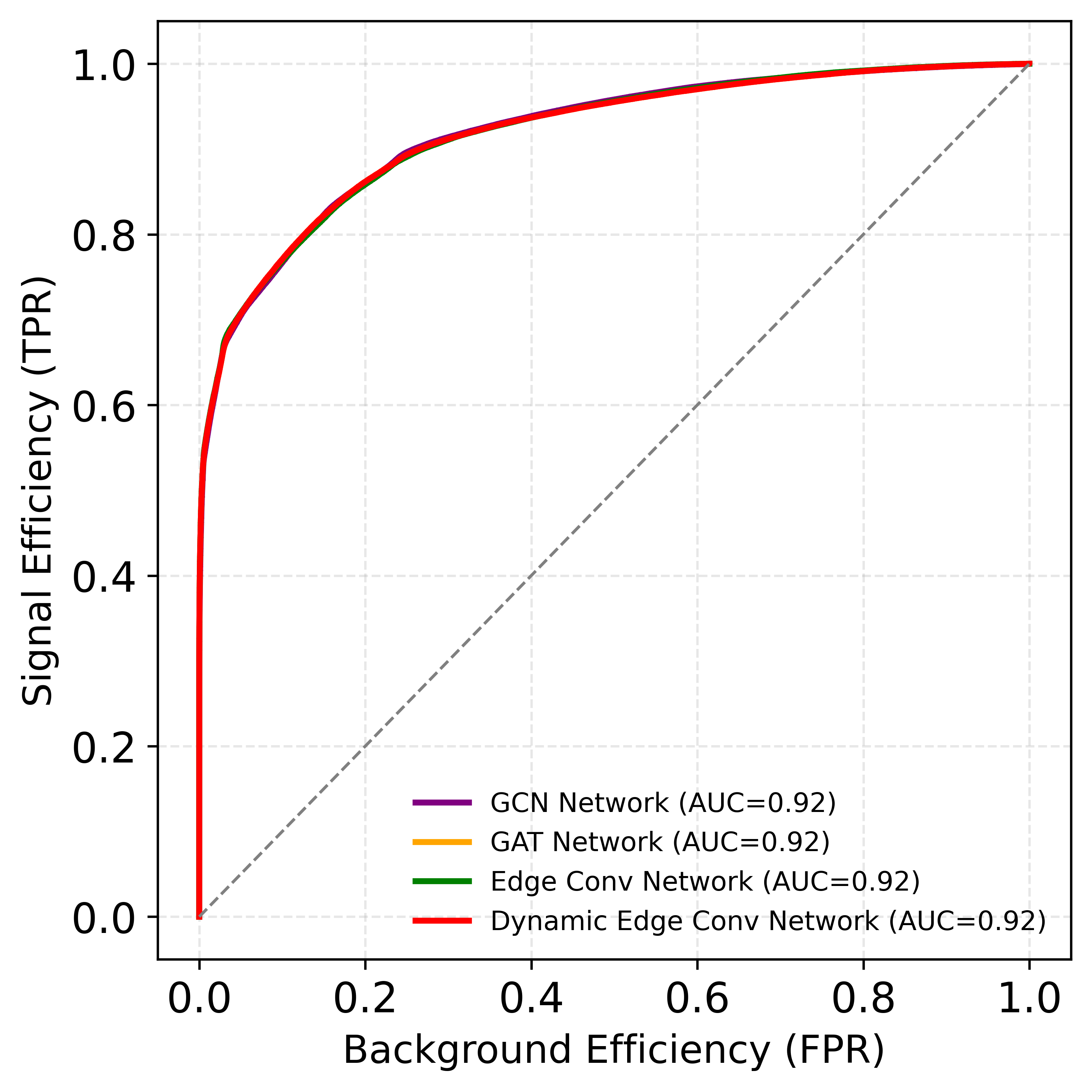}
\hfill
\includegraphics[width=0.32\textwidth]{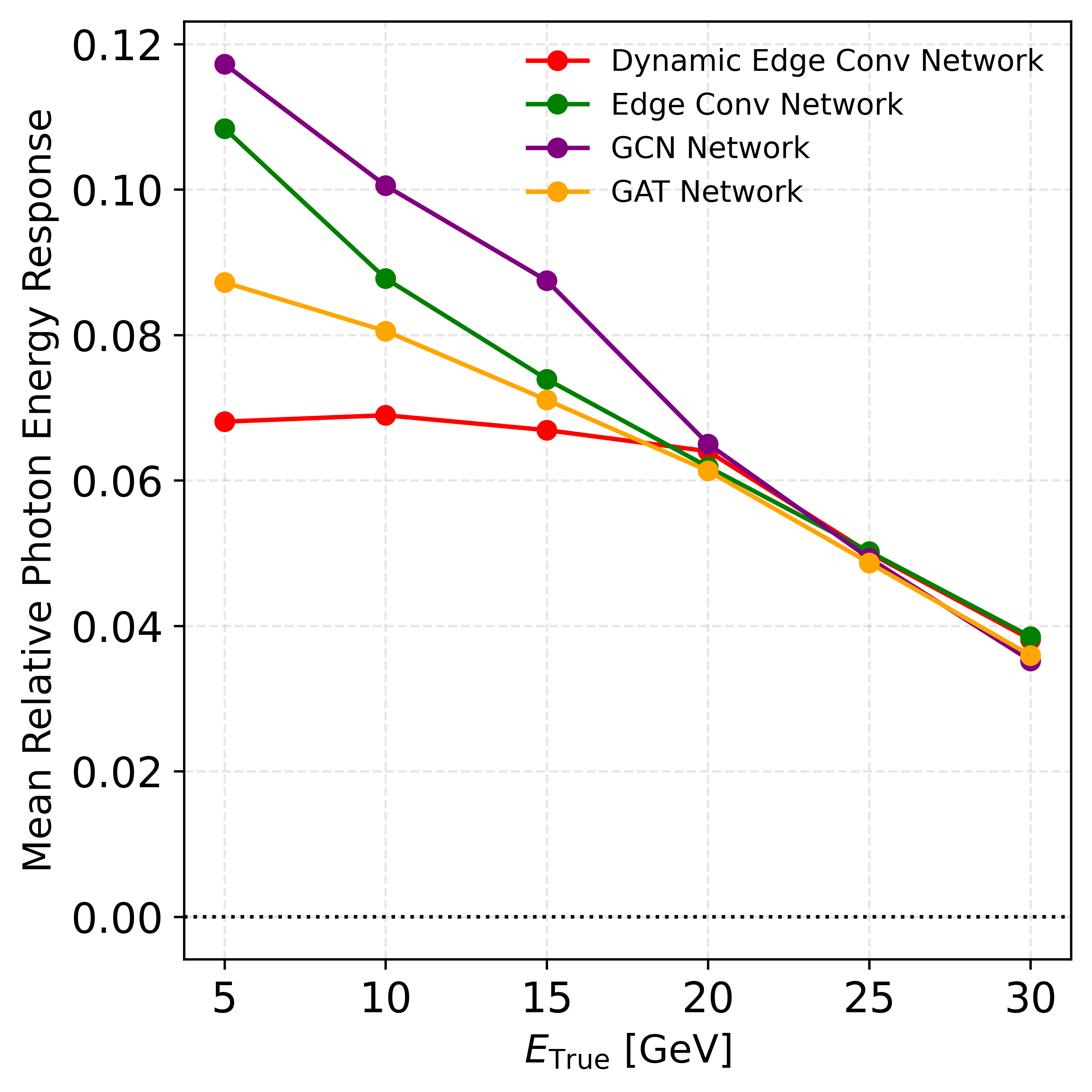}
\hfill
\includegraphics[width=0.32\textwidth]{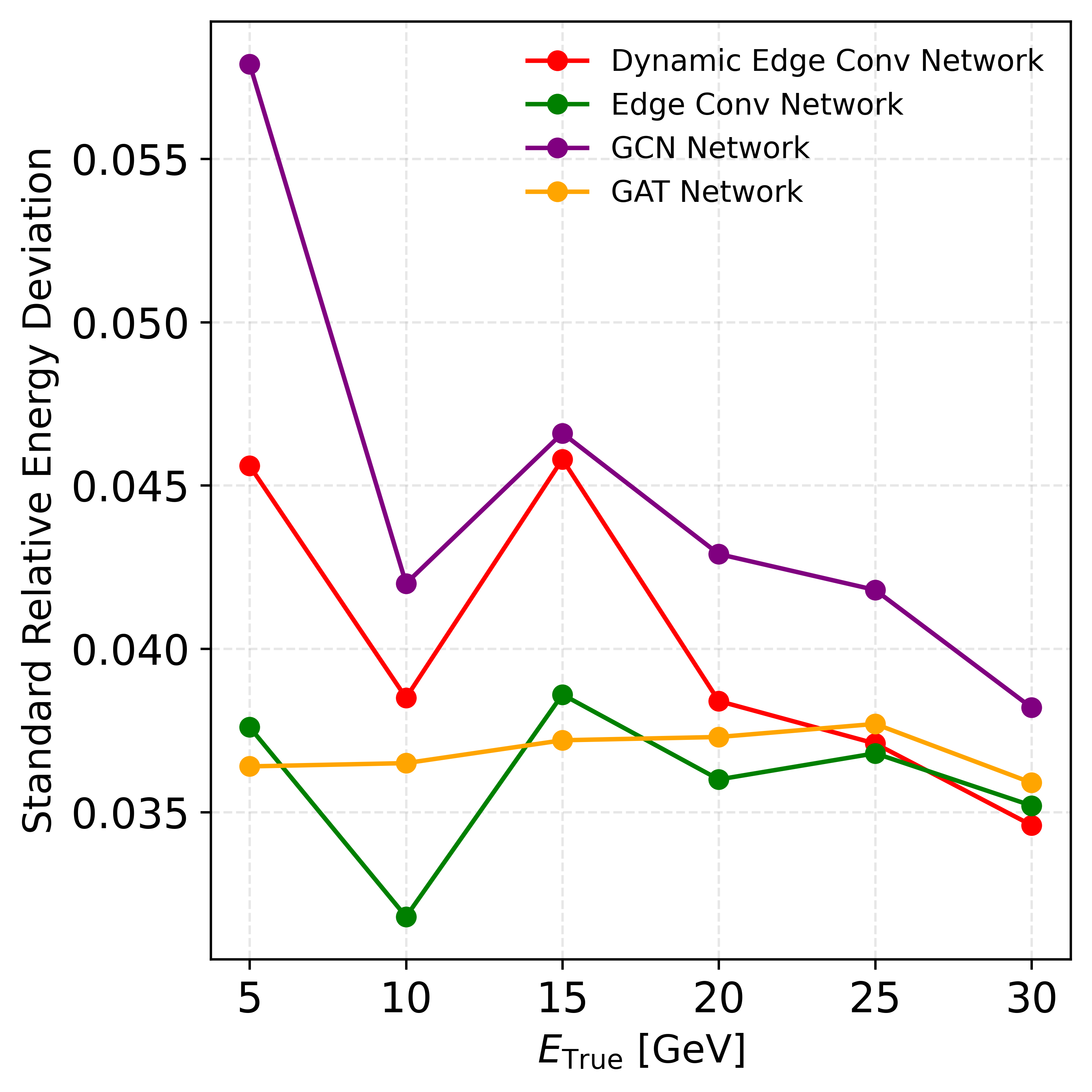}

\caption{Performance comparison of different GNN architectures. Left: ROC curves for different models, showing comparable discrimination power with nearly identical AUC values. Middle: Mean relative response of the reconstructed photon energy as a function of the true photon energy. Right: Standard deviation of the relative error on the reconstructed photon energy versus the true photon energy.}
\label{fig:ML_score_roc_comparison_models}
\end{figure*}

Figure~\ref{fig:ML_score_roc_comparison_models} (left) shows the ROC curves for the different GNN models. All architectures yield nearly identical ROC curves and AUC values, which indicates a comparable ability to discriminate signal hits from noise hits. However, the models show distinct performance in energy reconstruction. 
Figure~\ref{fig:ML_score_roc_comparison_models} (middle) displays the mean
relative response of the reconstructed photon energy as a function of true photon
energy. Both the GAT network and the Dynamic Edge Convolution network demonstrate
closer agreement with true photon energy across the full energy range compared to the other
architectures. But Dynamic Edge Convolution model performs less favorably when the resolution is considered. Figure~\ref{fig:ML_score_roc_comparison_models} (right) presents the standard deviation of the relative energy error. Here, the GAT consistently exhibits a lower and more uniform standard deviation across the entire evaluated energy range compared to the other model, demonstrating a superior and more stable energy resolution.

\begin{figure*}[htbp]
\centering
\includegraphics[width=0.44\textwidth]{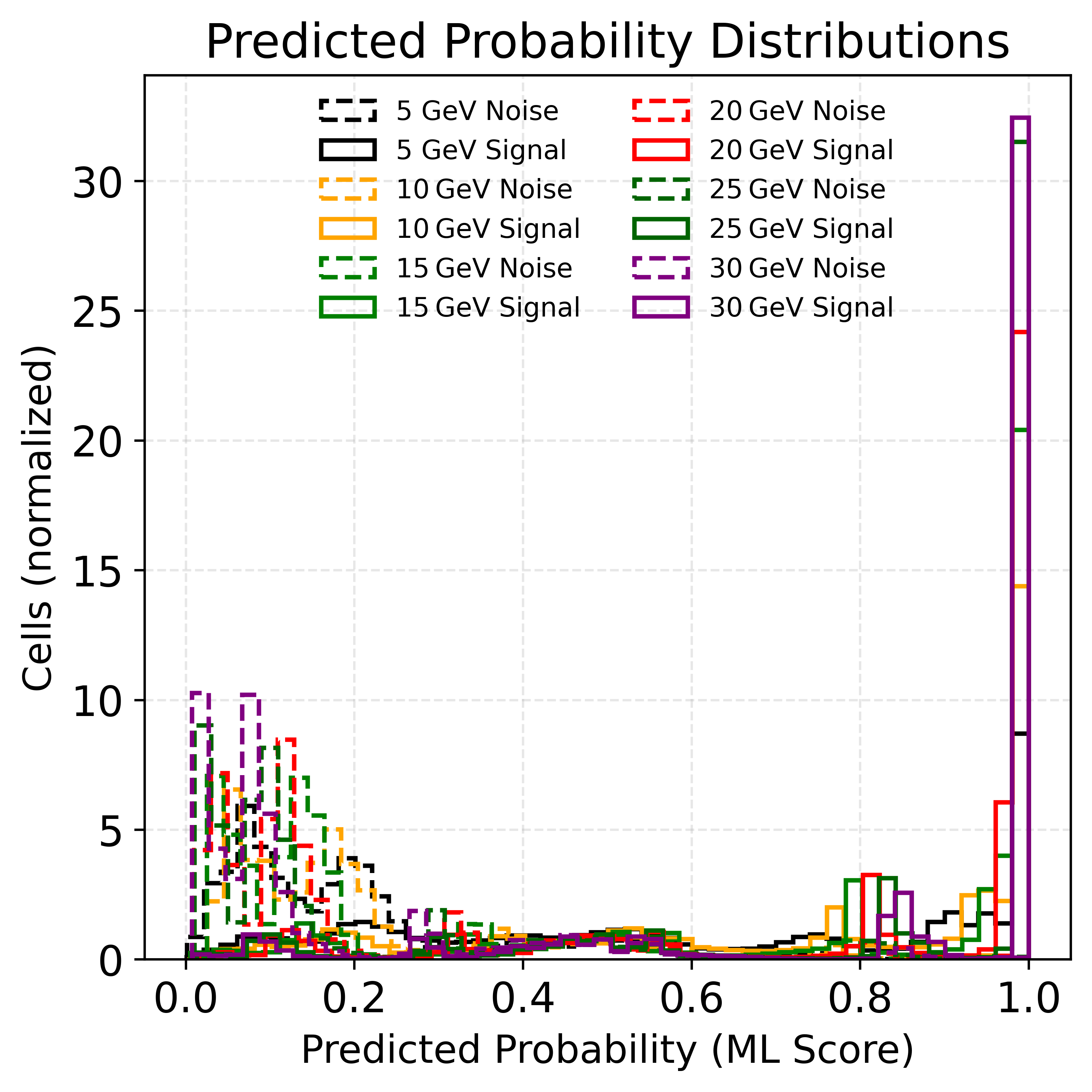}
\hfill
\includegraphics[width=0.44\textwidth]{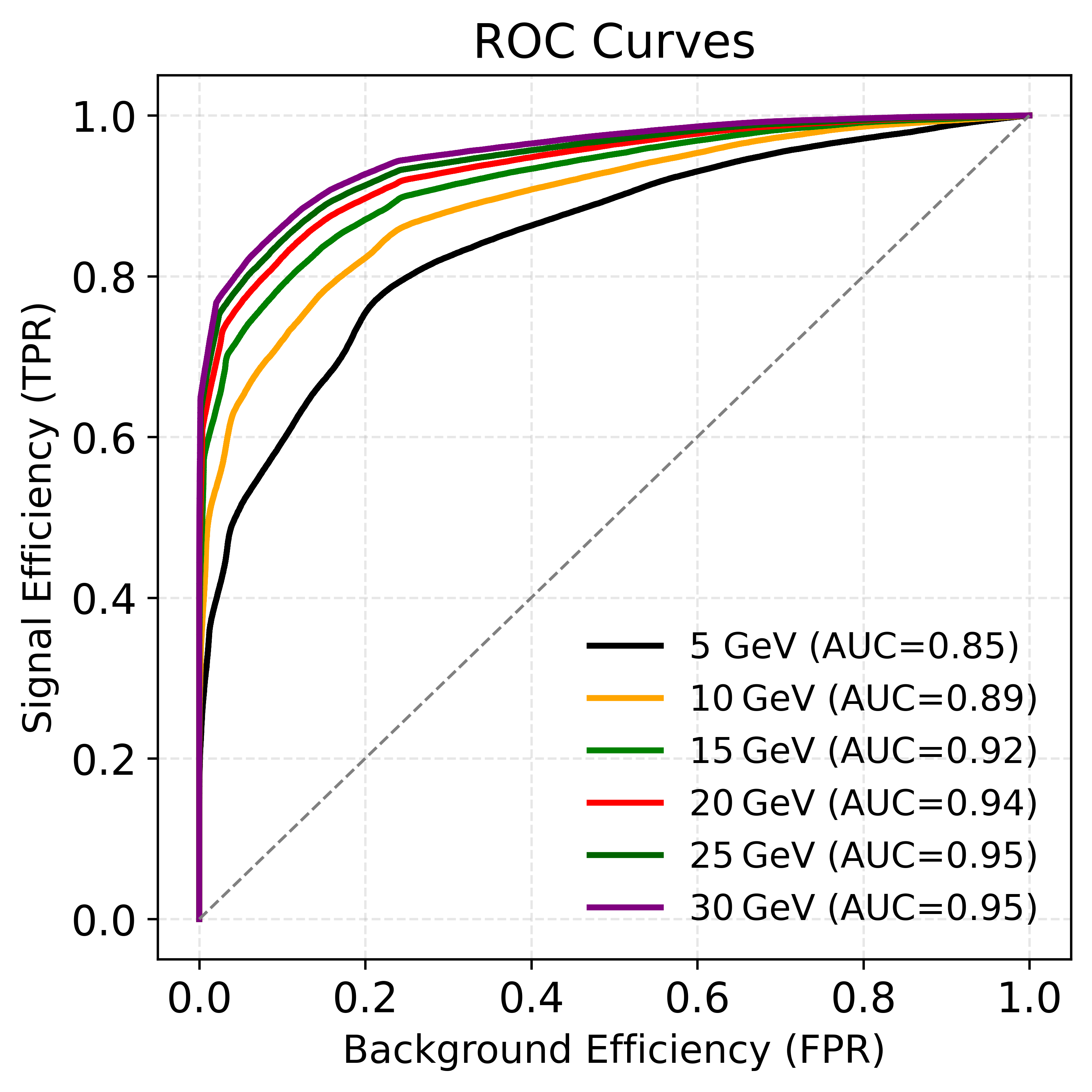}
\caption{Performance of the GAT network for different photon energies in discriminating signal cells from noise cells. Left: ML score distributions for signal and noise cells at various photon energies. Right: ROC curves with corresponding AUC values for different photon energies.}
\label{fig:ML_score_roc_comparison}
\end{figure*}

Based on these studies, we observe that the GAT-based GNN achieves the best performance, as it exploits local geometric structure using fixed detector adjacencies and relative features between neighboring hits, leading to more stable and accurate energy reconstruction. Therefore, it is selected as the optimal model for further analysis. Figure~\ref{fig:ML_score_roc_comparison} presents the ML score distributions for signal and noise cells and the corresponding ROC curves for different photon energies for this optimal model. The model performed consistently well with good signal and background separation for different photon energies. The selected model comprises approximately
$11.3 \times 10^{3}$ trainable parameters, was trained on a single NVIDIA Tesla
V100-PCIE-16GB GPU over 1500 epochs (approximately 8 hours), and achieves a mean
inference time of $4.38\,\mathrm{ms}$ per event.

In the second step, the reconstructed photon energy, \( E_{ \text{photon}}^{\text{reco}} \), is determined by summing the energies of the constituent cells, \( E_{i,\text{cell}}^{\text{reco}} \), that passed specific selection criteria. Two selection methods are evaluated: a simple threshold on the cell energy (`EnergyCut`) and a cut on an ML model score. The photon energy scale and resolution are determined from the mean and the standard deviation of the reconstructed photon energy distribution.

\begin{figure*}[htbp]
\centering
\includegraphics[width=0.32\textwidth]{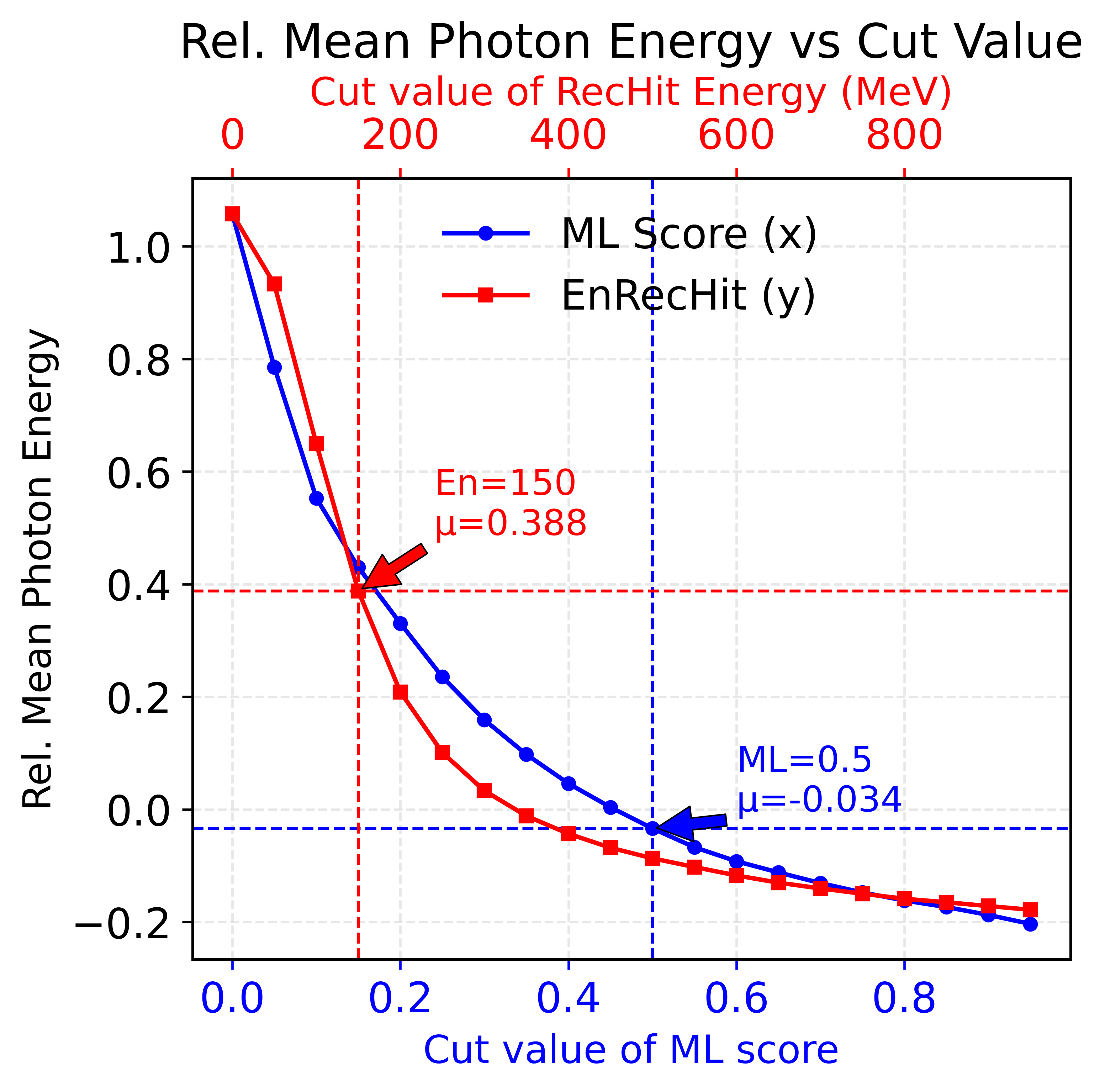}
\hfill
\includegraphics[width=0.32\textwidth]{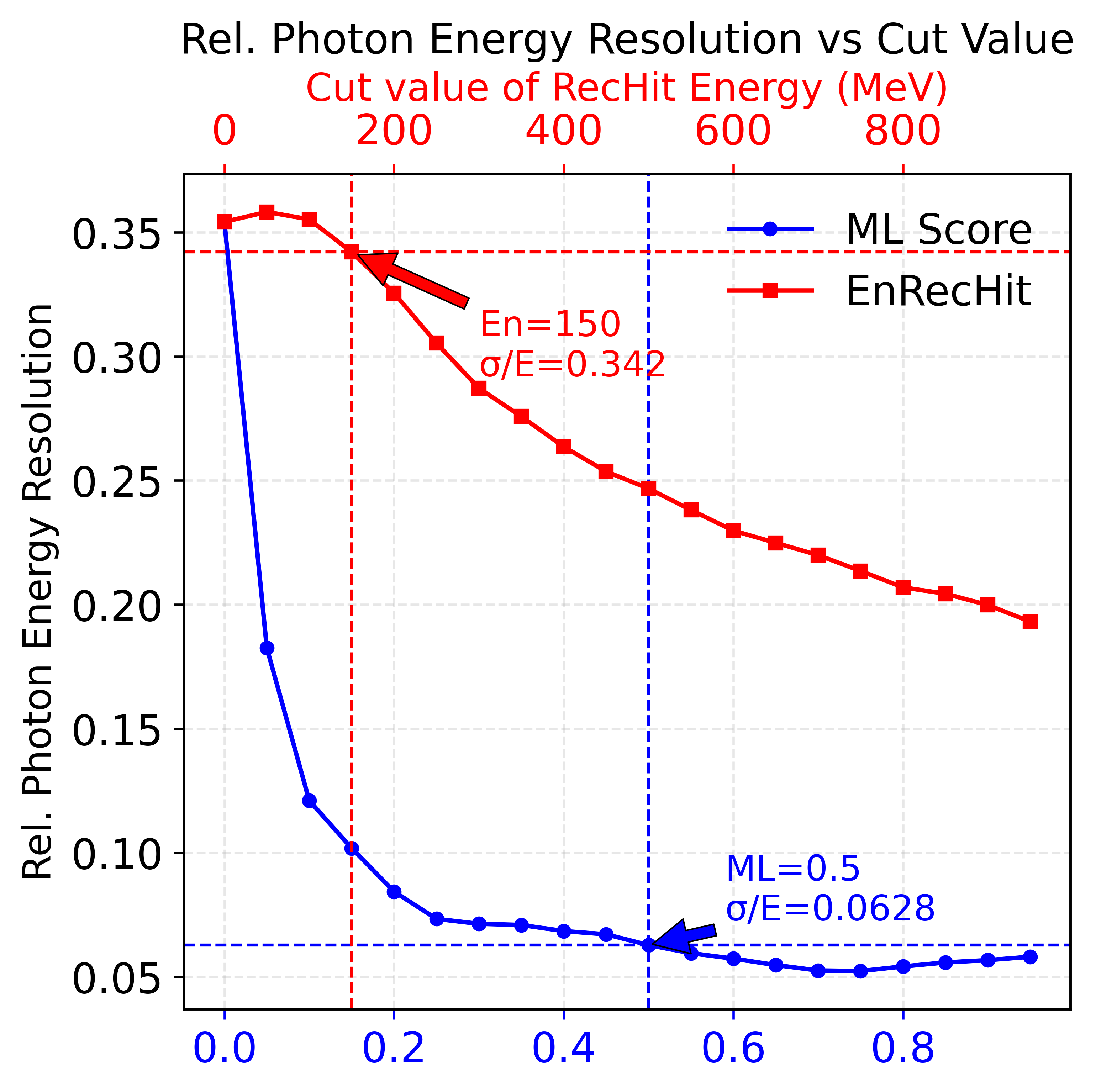}
\hfill
\includegraphics[width=0.32\textwidth]{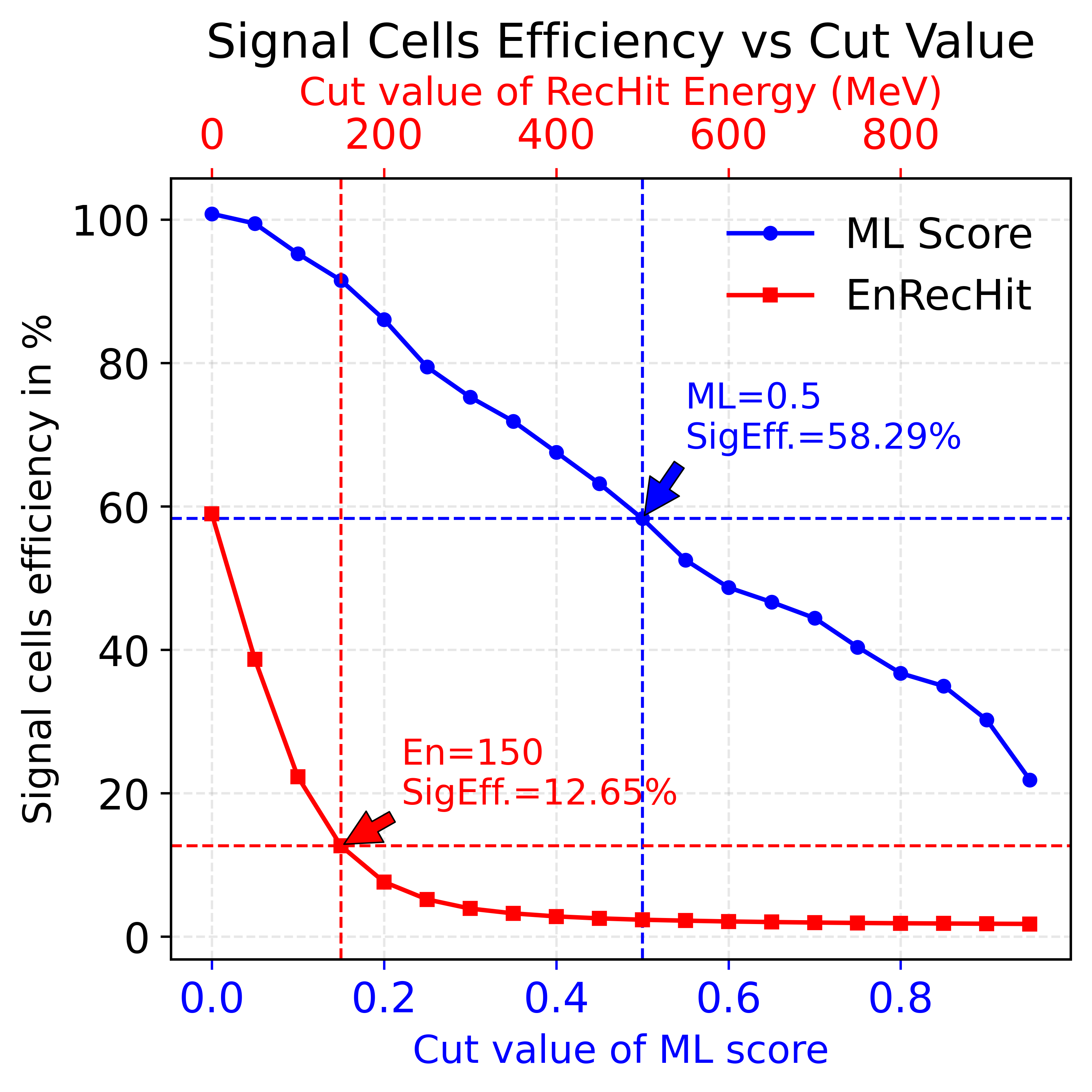}

\caption{Variation of relative mean of photon energy (left), relative photon energy resolution (middle), and signal cell reconstruction efficiency (right) as functions of the applied Energy Cut and ML score for 5~GeV photons.}
\label{fig:photon_metrics_true_5GeV}
\end{figure*}

\begin{figure*}[htbp]
\centering
\includegraphics[width=0.32\textwidth]{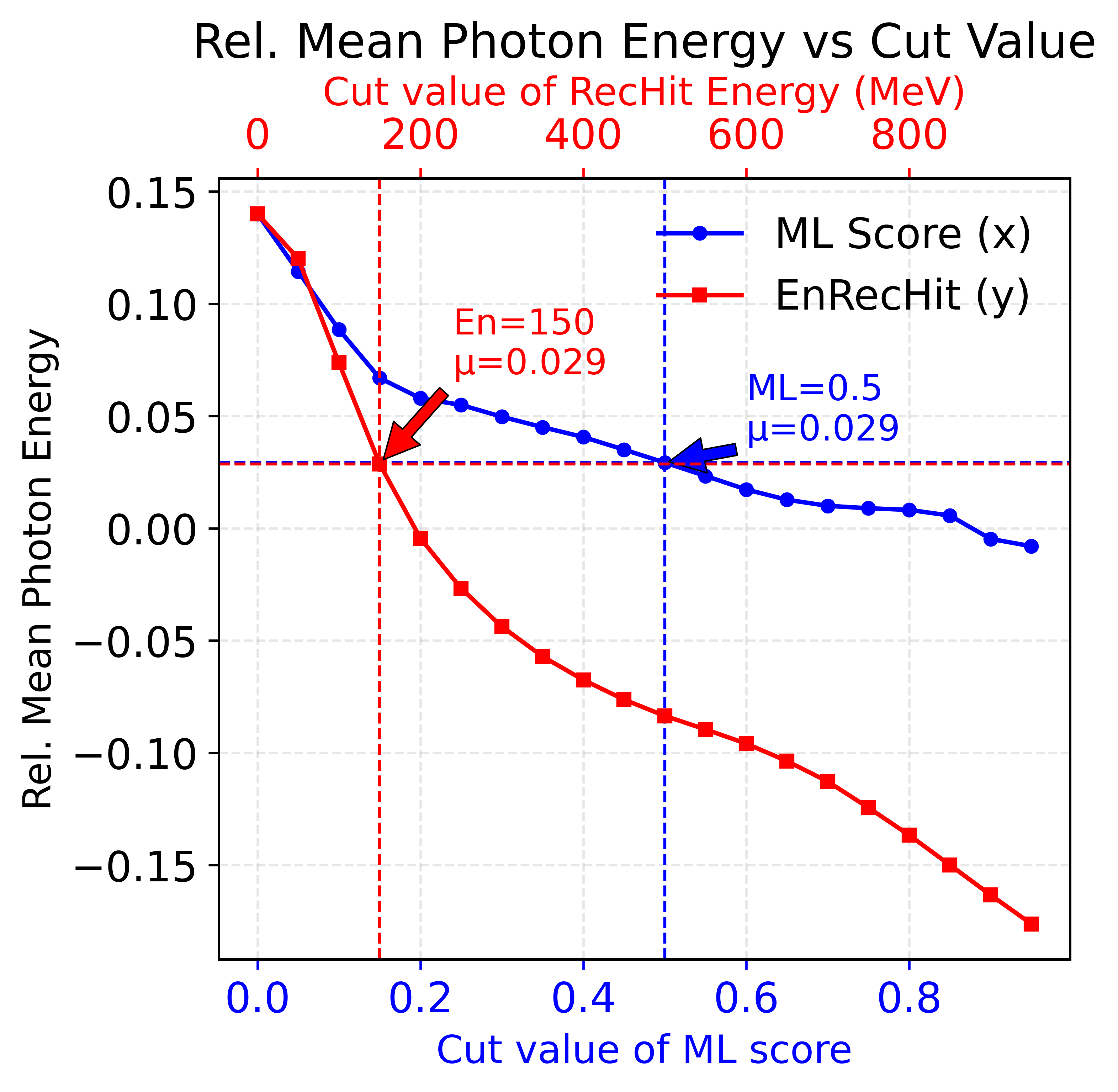}
\hfill
\includegraphics[width=0.32\textwidth]{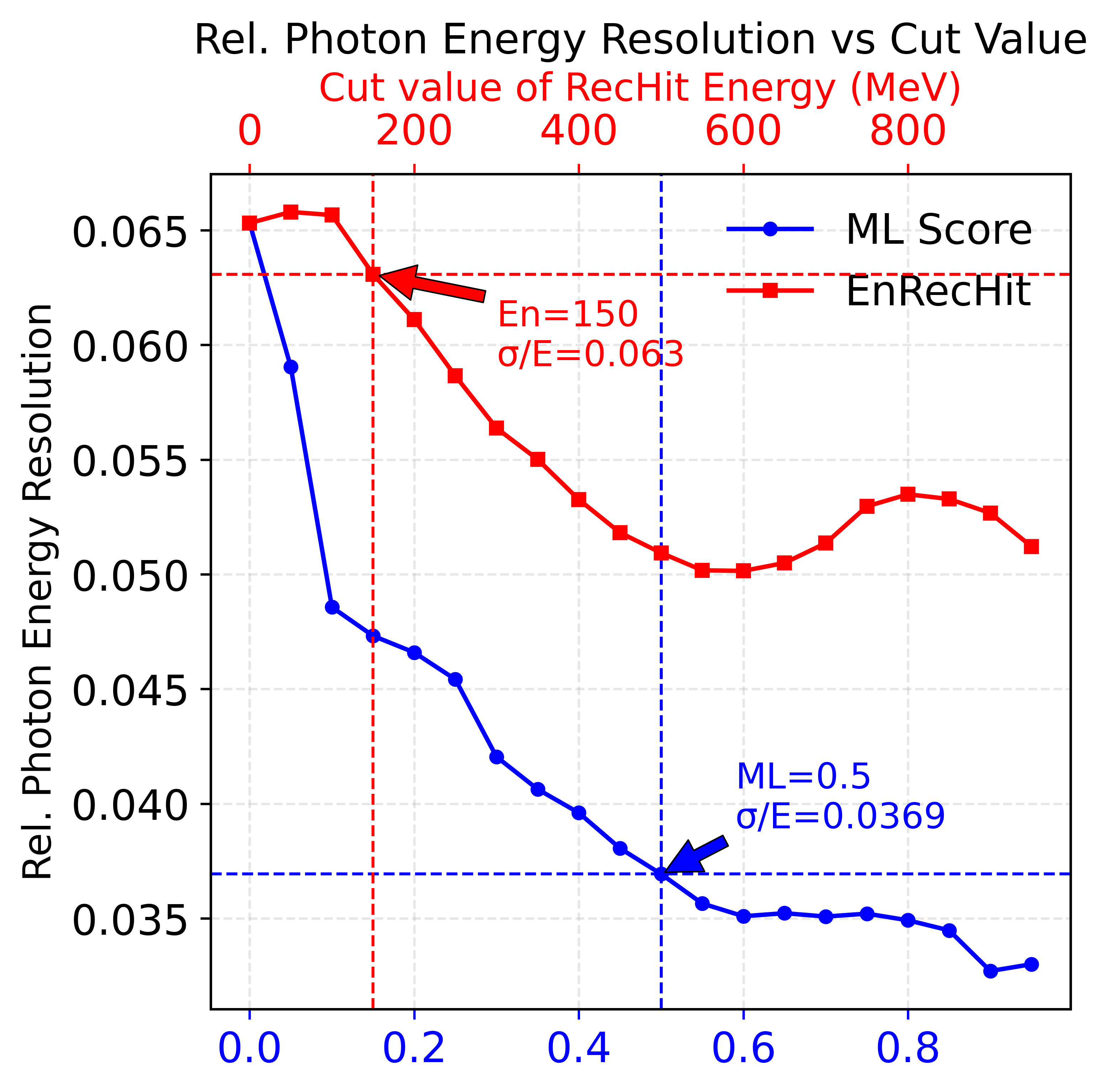}
\hfill
\includegraphics[width=0.32\textwidth]{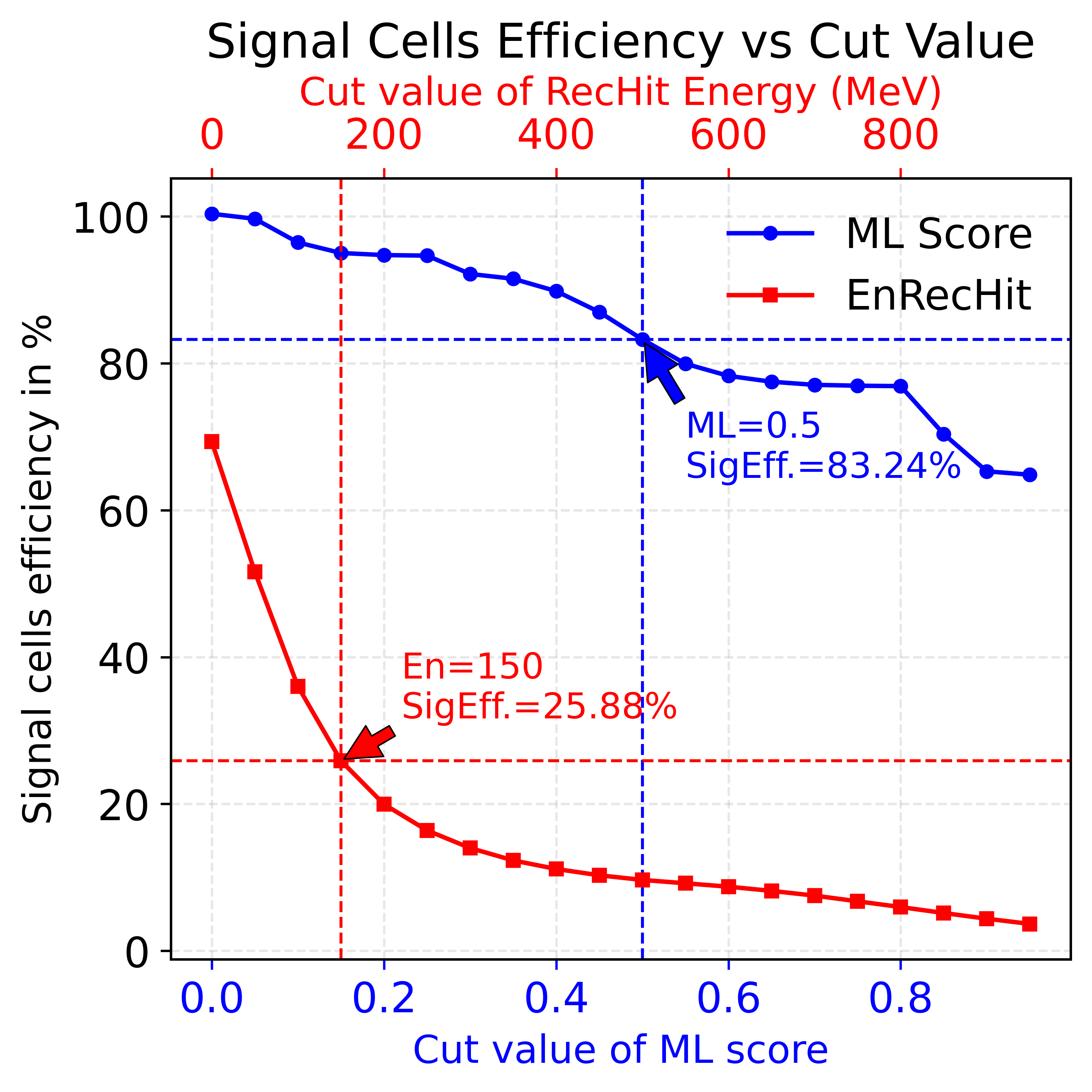}

\caption{Variation of relative mean of photon energy (left), relative photon energy resolution (middle), and signal cell reconstruction efficiency (right) as functions of the applied Energy Cut and ML score for 30~GeV photons.}
\label{fig:photon_metrics_true_30GeV}
\end{figure*}

To optimize the selection criteria, these performance metrics were studied as a function of the applied thresholds. Figures~\ref{fig:photon_metrics_true_5GeV} and~\ref{fig:photon_metrics_true_30GeV} show the relative mean photon energy (left) and relative energy resolution (middle) for 5 GeV and 30 GeV photons, respectively. These figures also show the signal cell efficiency (right), defined as the fraction of true signal cells retained after the selection cuts. The ML-based selection consistently achieves superior mean reconstructed photon energy and resolution, and greater signal cell efficiency compared to the energy-based cut—a level of performance that could not be replicated by any `EnergyCut` value. Based on this study, the ML score threshold was set to 0.5 to optimize these three metrics. For the baseline comparison, the `EnergyCut` threshold was set to 150 MeV, a value comparable to the simulated noise floor. 

\begin{figure*}[htbp]
\centering

\begin{subfigure}[b]{0.48\textwidth}
    \centering
    \includegraphics[width=\textwidth]{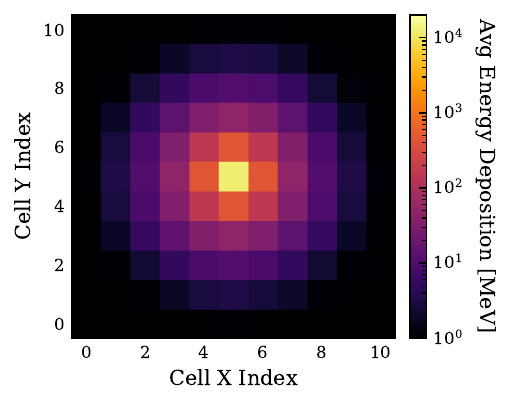}
    \label{fig:wo_true_energy}
\end{subfigure}
\hfill
\begin{subfigure}[b]{0.48\textwidth}
    \centering
    \includegraphics[width=\textwidth]{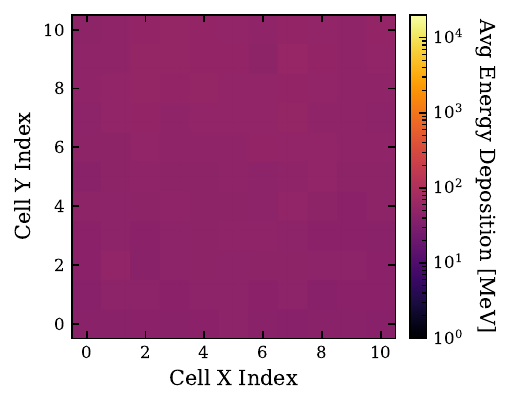}
    \label{fig:wo_pileup}
\end{subfigure}

\vspace{0.5em}

\begin{subfigure}[b]{0.48\textwidth}
    \centering
    \includegraphics[width=\textwidth]{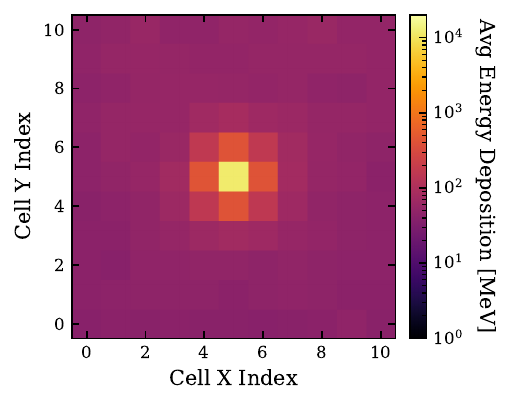}
    \label{fig:wo_cut_energy}
\end{subfigure}
\hfill
\begin{subfigure}[b]{0.48\textwidth}
    \centering
    \includegraphics[width=\textwidth]{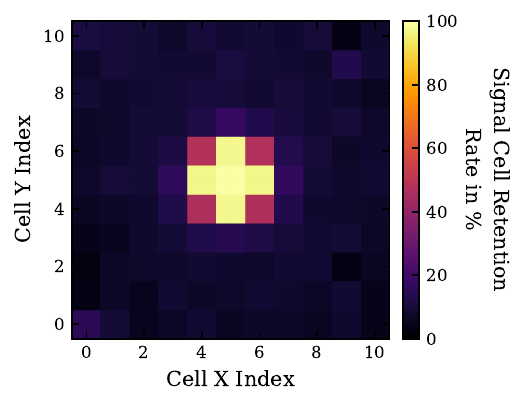}
    \label{fig:signal_efficiency_energycut}
\end{subfigure}

\caption{Average energy deposition and signal cell loss in an $11\times11$
ECAL window surrounding a 15~GeV photon, evaluated over simulated events. Top Left: True energy deposited by the
photon shower. Top Right: Contribution from pileup and electronic noise in the absence of a photon. Bottom Left: Reconstructed cell energy including the photon
signal, pileup, and electronic noise. Bottom Right: Signal cell retention rate obtained using "EnergyCut" method}
\label{fig:hitmap_pileup_smearing}
\end{figure*}

Figure~\ref{fig:hitmap_pileup_smearing} shows the average energy distributions of the true signal (top left) and pileup (top right) deposited in each of the $11\times11$ cells surrounding a 15~GeV photon, along with the probability of retaining a signal cell after applying the simple energy threshold (bottom right). This loss of signal cells is what the ML algorithm aims to recover.

\begin{figure*}[htbp]
\centering
\includegraphics[width=0.32\textwidth]{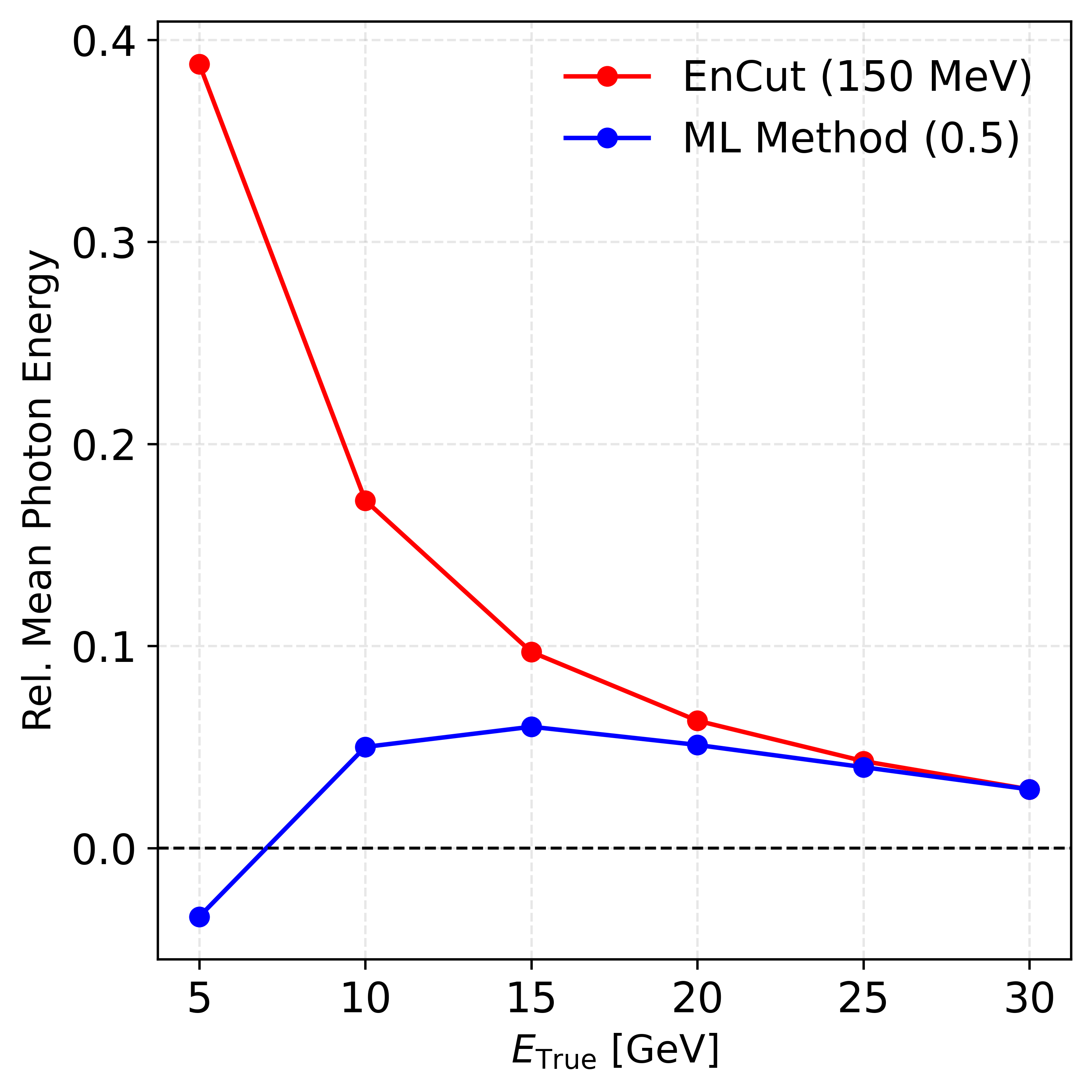}
\hfill
\includegraphics[width=0.32\textwidth]{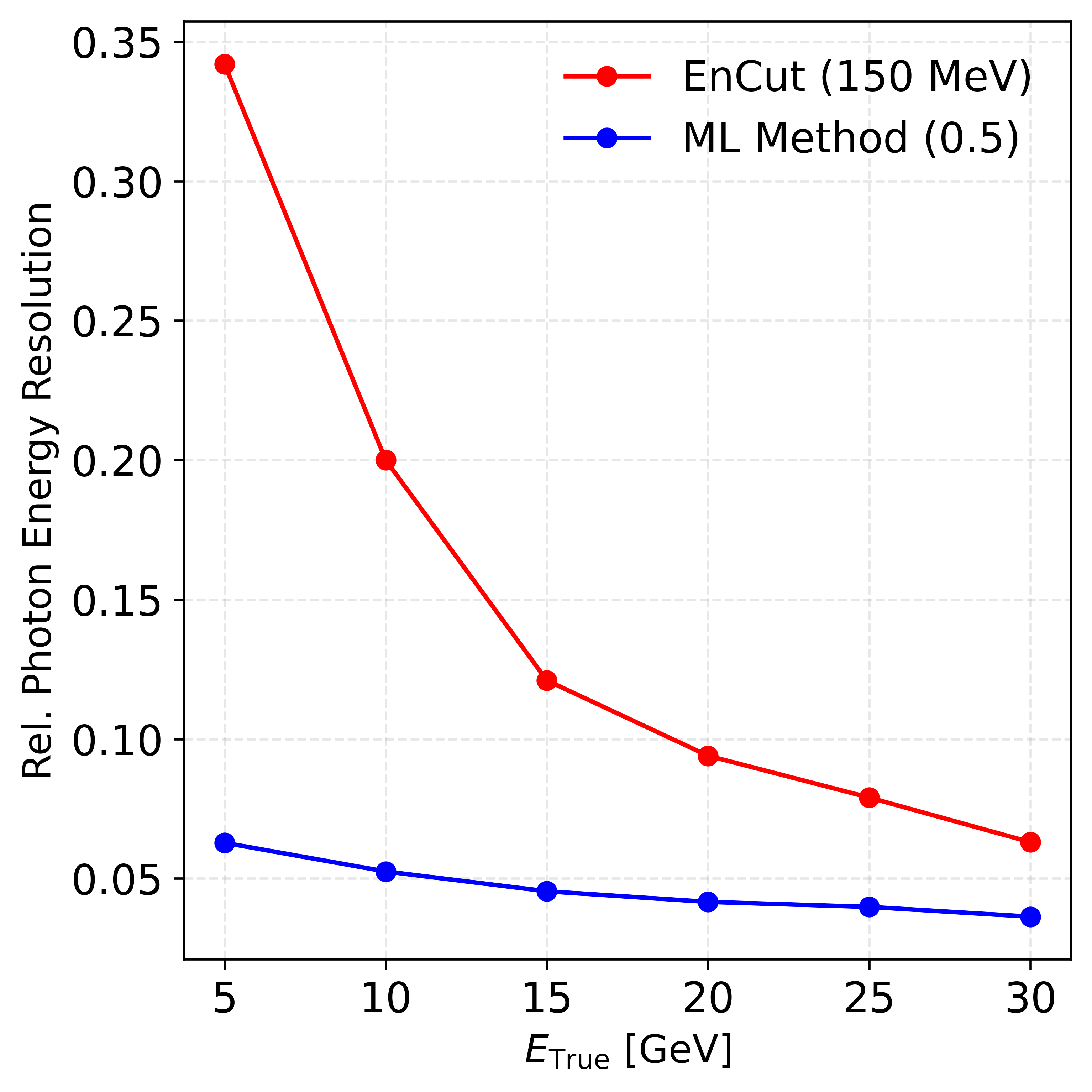}
\hfill
\includegraphics[width=0.32\textwidth]{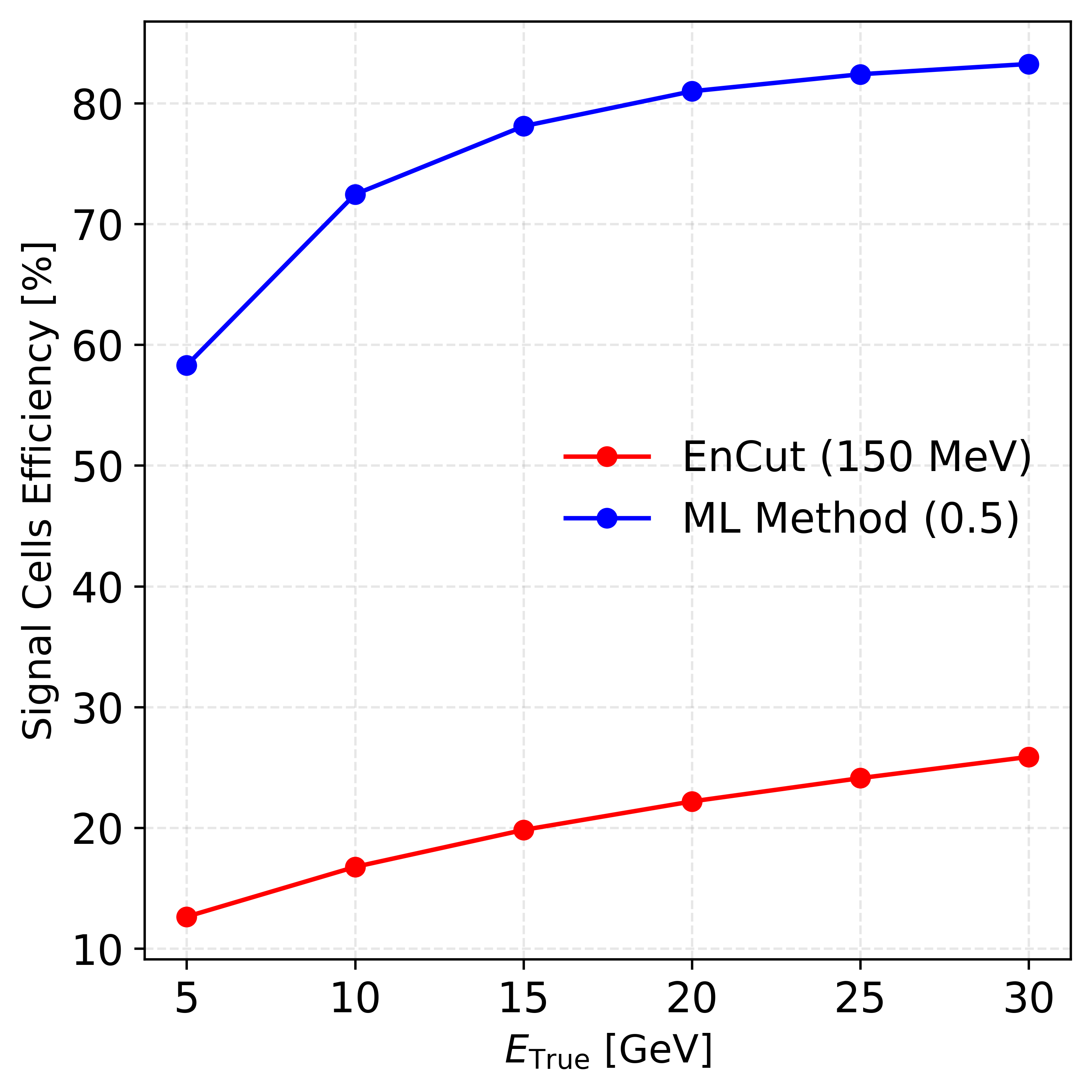}

\caption{Variation of relative mean of photon energy (left), relative photon energy resolution (middle), and signal cell reconstruction efficiency (right) as functions of photon energy at an optimized value of ML score and Energy Cut.}
\label{fig:photon_metrics_true_vs_energy}
\end{figure*}

Finally, Figure~\ref{fig:photon_metrics_true_vs_energy} shows the performance of both methods using these optimized thresholds as a function of the true photon energy. The ML-based approach demonstrates consistently superior energy reconstruction across all generated photon energy points.



\section{Conclusion}

Elimination of noisy calorimeter cells while preserving cells with genuine energy deposits can be a difficult optimization problem in calorimetric energy reconstruction, in particular in a high radiation environment where one expects large occupancy of the cells and also high rate of degradation of cells with time. 

In this work, we demonstrated how one can use machine learning (ML) methods to substantially improve the performance of calorimetric energy measurement under such high radiation environment. We developed a graph neural network (GNN) which can efficiently compensate for the degradation in energy resolution induced by irradiation of the electromagnetic calorimeter (ECAL) crystals and associated electronics, by exploiting the full pulse-shape information and correlation between individual cell energy deposits inside an electromagnetic cluster. 

The power of the GNN comes from its capability to dynamically discriminate between noisy hits and true hits without the need of a fixed energy threshold. 
We did this study using a standalone GEANT4 simulation of an 11$\times$11 matrix of lead-tungstate crystals. 
Compared to a simple threshold-based mitigation strategy, our GNN approach yields an improvement in energy resolution of up to $\sim82\%$ and recovers up to $\sim83\%$ of true signal cells within the calorimeter grid.


\clearpage

\raggedbottom


\begin{thebibliography}{99}



\bibitem{Evans:2008zzb}
L. Evans, P. Bryant, JINST \textbf{3}, S08001 (2008). 
\href{https://doi.org/10.1088/1748-0221/3/08/S08001}{doi:10.1088/1748-0221/3/08/S08001}
\bibitem{CMS:2024ppo}
CMS Collaboration, A. Hayrapetyan et al., JINST \textbf{19}, P09004 (2024). 
\href{https://doi.org/10.1088/1748-0221/19/09/P09004}{doi:10.1088/1748-0221/19/09/P09004}
\bibitem{Dutta_2023}
S. Dutta, S. Ghosh, S. Bhattacharya, S. Saha, JINST \textbf{18}, P03038 (2023). 
\href{https://doi.org/10.1088/1748-0221/18/03/P03038}{doi:10.1088/1748-0221/18/03/P03038}
\bibitem{GEANT4}
GEANT4 Collaboration, S. Agostinelli et al., Nucl. Instrum. Meth. A \textbf{506}, 250 (2003). 
\href{https://doi.org/10.1016/S0168-9002(03)01368-8}{doi:10.1016/S0168-9002(03)01368-8}
\bibitem{CMS:2008xjf}
CMS Collaboration, S. Chatrchyan et al., JINST \textbf{3}, S08004 (2008). 
\href{https://doi.org/10.1088/1748-0221/3/08/S08004}{doi:10.1088/1748-0221/3/08/S08004}
\bibitem{CMS:1997ysd}
CMS Collaboration, \textit{The CMS electromagnetic calorimeter project: Technical Design Report}, CERN-LHCC-97-033 (1997)
\bibitem{CMS:2020xlg}
CMS Collaboration, A.M. Sirunyan et al., JINST \textbf{15}, P10002 (2020). 
\href{https://doi.org/10.1088/1748-0221/15/10/P10002}{doi:10.1088/1748-0221/15/10/P10002}
\bibitem{Adzic:2006hda}
P. Adzic et al., Eur. Phys. J. C \textbf{46}, 23 (2006). 
\href{https://doi.org/10.1140/epjcd/s2006-02-002-x}{doi:10.1140/epjcd/s2006-02-002-x}


\bibitem{pythia8}
T. Sj\"ostrand et al., Comput. Phys. Commun. \textbf{191}, 159 (2015). 
\href{https://doi.org/10.1016/j.cpc.2015.01.024}{doi:10.1016/j.cpc.2015.01.024}


\bibitem{BUCKLEY2021107310}
A. Buckley et al., Comput. Phys. Commun. \textbf{260}, 107310 (2021). 
\href{https://doi.org/10.1016/j.cpc.2020.107310}{doi:10.1016/j.cpc.2020.107310}


\bibitem{Hastie:2009esl}
T. Hastie, R. Tibshirani, J. Friedman, \textit{The Elements of Statistical Learning}, 2nd edn. (Springer, New York, 2009)


\bibitem{Krizhevsky:2012alexnet}
A. Krizhevsky, I. Sutskever, G.E. Hinton, in \textit{Advances in Neural Information Processing Systems 25} (2012)

\bibitem{Wu:2021gnn}
Z. Wu et al., IEEE Trans. Neural Netw. Learn. Syst. \textbf{32}, 4 (2021). 
\href{https://doi.org/10.1109/TNNLS.2020.2978386}{doi:10.1109/TNNLS.2020.2978386}

\bibitem{Kipf:2017gcn}
T.N. Kipf, M. Welling, in \textit{Proceedings of ICLR 2017} (2017). arXiv:1609.02907 [cs.LG]

\bibitem{Velickovic:2018gat}
P. Veličković et al., in \textit{Proceedings of ICLR 2018} (2018). arXiv:1710.10903 [stat.ML]


\bibitem{Wang:2019dgcnn}
Y. Wang et al., ACM Trans. Graph. \textbf{38}, 5 (2019). arXiv:1801.07829 [cs.CV]



\end{thebibliography}
\end{document}